\documentclass[10pt,twocolumn]{IEEEtran}

\IEEEoverridecommandlockouts

\usepackage{graphicx}
\usepackage{amsmath}
\usepackage{amssymb}
\usepackage[caption=false]{subfig}
\usepackage[noadjust]{cite}
\usepackage{float}
\usepackage{algorithm}
\usepackage{bibentry}
\usepackage{balance}
\usepackage{algorithm}
\usepackage{algorithmicx}
\usepackage{algpseudocode}
\usepackage{xcolor}
\usepackage{subfig}

\usepackage{url}

\graphicspath{{img/}}

\begin{document}

\bstctlcite{IEEEexample:BSTcontrol}

\title{Impact of 3D Antenna Radiation Pattern in UAV Air-to-Ground Path Loss Modeling and RSRP-based Localization in Rural Area\thanks{This research is supported in part by the NSF award CNS-1939334. The data and the Matlab scripts for generating the results in this manuscript are available at~\cite{IEEEDataPort_2}.}\thanks{S. J. Maeng, H. Kwon,  O. Ozdemir, and \.{I}. G\"{u}ven\c{c} are with the Department of Electrical and Computer Engineering, North Carolina State University, Raleigh, NC 27606 USA (e-mail: smaeng@ncsu.edu; khyeokj@ncsu.edu; oozdemi@ncsu.edu; iguvenc@ncsu.edu).}}

\author{\IEEEauthorblockN{Sung Joon Maeng, Hyeokjun Kwon, Ozgur Ozdemir, \textit{Member, IEEE}, AND \.{I}smail G\"{u}ven\c{c}, \textit{Fellow, IEEE}}}

\maketitle

\begin{abstract}
Ensuring reliable and seamless wireless connectivity for unmanned aerial vehicles (UAVs) has emerged as a critical requirement for a wide range of applications. The increasing deployment of UAVs has increased the significance of cellular-connected UAVs (C-UAVs) in enabling beyond-visual line of sight (BVLOS) communications. To ensure the successful operation of C-UAVs within existing terrestrial networks, it is vital to understand the distinctive characteristics associated with air-to-ground signal propagation. In this paper, we investigate the impact of 3D antenna patterns on a UAV air-to-ground path loss model, utilizing datasets obtained from a measurement campaign. We conducted UAV experiments in a rural area at various fixed heights, while also characterizing the 3D antenna radiation pattern by using an anechoic chamber facility. By analyzing reference signal received power (RSRP) using path loss models that account for antenna patterns, we observed that our measurement results, obtained at different UAV heights, aligned well with the two-ray path loss model when incorporating the measured antenna pattern. We propose an RSRP-based localization algorithm at a UAV that takes into account antenna patterns in both offline and online scenarios. Through our experimentation dataset, we show that incorporating measured antenna patterns significantly enhances the source localization accuracy.
\end{abstract}


\section{Introduction} \label{sec:intro}
Drones, also referred to as unmanned aerial vehicles (UAVs), have gained significant attention in recent years due to their wide range of promising applications. They are being increasingly utilized for various purposes in military, commercial, and public safety areas including surveillance~\cite{namuduri2022advanced}, delivery services~\cite{Insider_Intelligence}, drone taxi services~\cite{yun2021distributed}, live video streaming, and search and rescue~\cite{erdos2013experimental}. However, to fully realize the potential of these future applications, it is crucial to establish wireless connectivity with UAVs in beyond-visual-line-of-sight (BVLOS) scenarios. This can be achieved through the use of cellular-connected UAVs (C-UAVs), which enable seamless communication and data exchange between the UAVs and the ground base station (BS)~\cite{zeng2018cellular}.

To successfully operate C-UAVs using the existing terrestrial networks, it is crucial to have a comprehensive understanding of the unique characteristics associated with air-to-ground signal propagation. In particular, two key factors contribute to the radio propagation model for C-UAVs: 1) elevation angle-dependent antenna radiation patterns and 2) strong ground reflection.  
%which have relatively modest effects in conventional terrestrial networks. 
In ground-to-ground wireless links, the altitude of the transmitter (Tx) and receiver (Rx) are often assumed fixed and relatively similar, and hence the corresponding radio propagation can often be reasonably modeled in the 2D space. However, as 3D geometry is considered in the air-to-ground C-UAV scenarios, the effects of the 3D antenna patterns and the ground reflection should be investigated more closely. 

In addition to their effects on wireless coverage, air-to-ground propagation characteristics for C-UAVs also critically affect 3D wireless localization accuracy with C-UAVs. Accurate and rapid localization of signal sources with UAVs has various use cases in commercial, public safety, and military scenarios. For instance, a UAV can be deployed for search and rescue of victims during/after natural disasters~\cite{atif2021uav}, or they can surveil enemy target locations to gather crucial information in a battlefield~\cite{jee2017autonomous}. 3D localization accuracy with UAVs can be improved significantly if the UAV-to-ground signal propagation and the 3D radiation patterns of transmit and receive antennas can be modeled accurately.

There have been several recent works in the literature focusing on UAV air-to-ground propagation models considering the impact of ground reflection and 3D antenna radiation patterns. In \cite{matolak2015unmanned}, air-to-ground channel characteristics have been modeled using flight measurement datasets. The research demonstrated that the two-ray path loss model, which incorporates ground reflection, helped to more accurately model air-to-ground signal measurements. In \cite{khawaja2020ultra}, UAV air-to-ground channel has been measured in the open rural area employing various antenna orientation setups. The findings revealed a substantial dependence of the received signal power on the elevation-domain antenna pattern. In \cite{al2017modeling}, the excess path loss in the cellular-to-UAV channel models due to the effect of the 3D antenna pattern has been analyzed. In \cite{badi2020experimentally}, the impact of a UAV's body and the placement and orientation of an antenna on the azimuth antenna pattern and polarization characteristics has been investigated while considering an anechoic chamber for antenna pattern measurement and field measurement conducted through both UAV-to-UAV and ground-to-UAV experimental setups.

In \cite{sinha2022impact}, considering the measurements at a terrestrial wireless network, the impact of 3D antenna radiation patterns on time difference of arrival (TDOA)-based 3D localization of UAVs has been studied, which reveals that antenna patterns significantly influence localization accuracy.

In this paper, first, we investigate the impact of 3D antenna patterns and ground reflections on air-to-ground C-UAV path loss models. We conducted the experiments in the National Science Foundation (NSF) Aerial Experimentation and Research Platform on Advanced Wireless (AERPAW) testbed site~\cite{9061157}. We flew the UAV at various heights and recorded I/Q samples at a software-defined radio (SDR) carried by the UAV corresponding to long-term evolution (LTE) signals transmitted by a BS on the ground. Following the experiments, we post-process the collected LTE I/Q samples dataset and obtain the reference signal received power (RSRP) at different UAV locations. The main contributions of this manuscript can be summarized as follows. 
\begin{itemize}
\item We analytically model air-to-ground radio propagation using different path loss models that explicitly take into account the measured antenna patterns and the ground reflection. By comparing the received signal measurements with the path loss models derived analytically, we show that the 3D antenna pattern is a critical factor in developing an accurate air-to-ground channel model. %Furthermore, through a comparison between the results obtained from the two-ray path loss model and the free space path loss model, we highlight the significance of including the ground reflection path in achieving an accurate air-to-ground path loss model. 
\item We propose an RSRP-based ground signal source localization algorithm with the UAV considering the impact of transmitter and receiver antenna patterns. We apply the algorithm in localizing the LTE source BS using the real-world measurements collected across the trajectory of the UAV. 
\item We evaluate the performance of both offline (at the end of the UAV flight) and online (real-time as the UAV flies) localization techniques. Simulation results using real-world LTE measurements show that the use of accurate antenna pattern models can help significantly to improve source localization accuracy using a UAV.
\end{itemize}

This paper is organized as follows. In Section~\ref{Sec:MeasurementCampaign}, we describe the measurement campaign for the UAV experiment and post-processing procedures of the collected LTE I/Q samples. In Section~\ref{Sec:A2G_propagation_model}, we present our air-to-ground propagation models that take into account ground reflections and 3D antenna radiation patterns. In Section~\ref{Sec:anal_ant_pat_a2g}, we compare the measured RSRP from the experiments, with the RSRP obtained using our proposed air-to-ground propagation models that take into account different antenna patterns. In Section~\ref{sec:localization}, we propose and discuss RSRP-based ground signal source localization techniques for both offline and online scenarios. In Section~\ref{Sec:num_results_localization}, we evaluate the accuracy of our proposed UAV-based source localization techniques using real-world datasets, and finally, the last section concludes the paper.
\begin{figure}[t!]
	\centering
	\subfloat[The AERPAW LWRFL site where UAV air-to-ground propagation data has been collected.]{\includegraphics[width=0.48\textwidth]{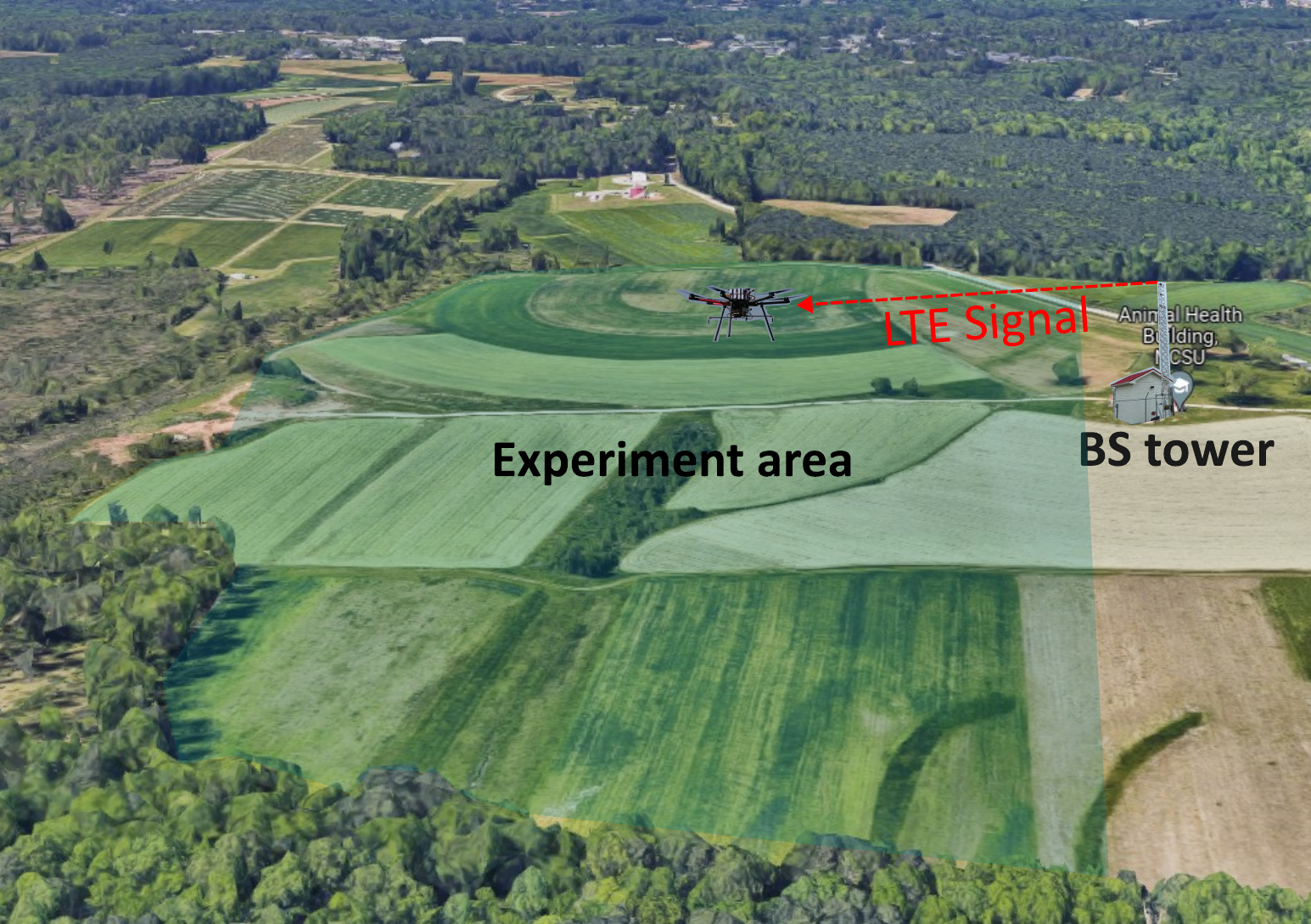}\label{fig:experiment_site}}
 
        \subfloat[The antenna installment of the UAV where the downward Rx antenna is deployed at the bottom of the UAV.]{\includegraphics[width=0.48\textwidth]{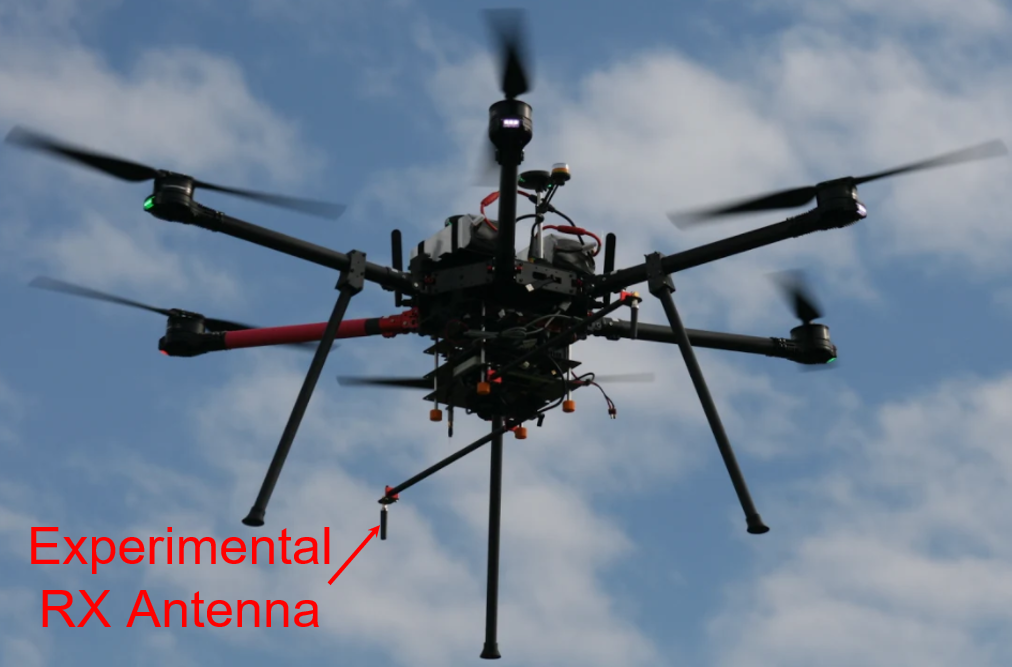}\label{fig:drone}}
	\caption{The illustration of the experiment site and the antenna setup of the UAV.}
\end{figure}
\begin{figure}[t!]
	\centering
	\includegraphics[width=0.48\textwidth]{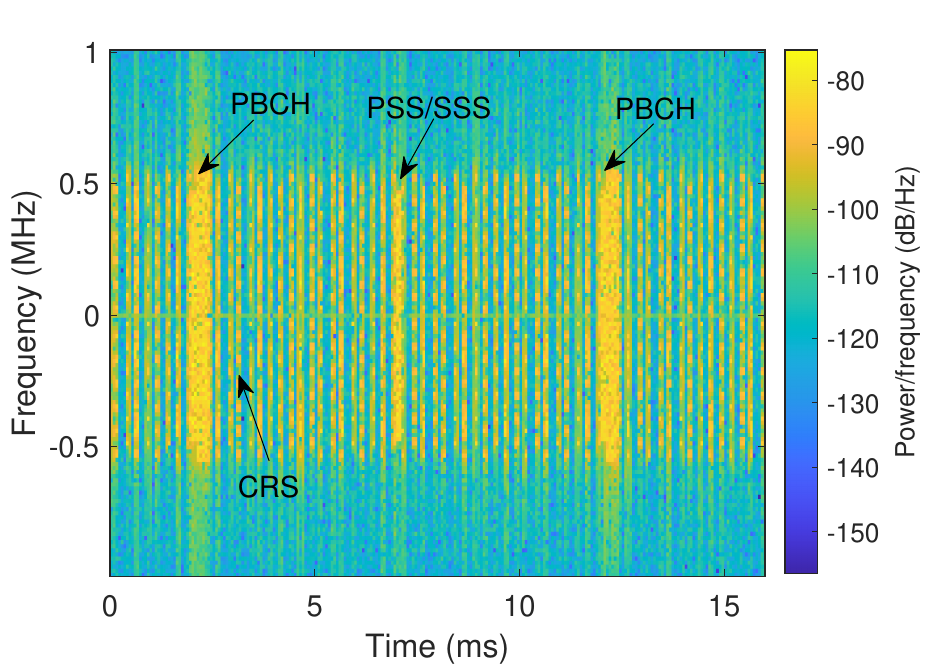}
	\caption{The spectrogram of the collected I/Q samples during the UAV flight shows the received LTE signal strength and noise plus interference floor level at the UAV receiver. In the transmitted LTE signal, the PBCH, the CRS, and the PSS/SSS are allocated.}\label{fig:spectro}
\end{figure}

\section{Measurement Campaign}\label{Sec:MeasurementCampaign}
In this section, we describe how we conduct the UAV and I/Q data collection experiments at NSF AERPAW Lake Wheeler Road Field Labs (LWRFL) site in Raleigh NC, USA.

\subsection{UAV Experiment Setup}
In our experiment, a BS tower transmits the LTE signal using the srsRAN open-source software-defined radio (SDR) software. We deploy a UAV that follows preprogrammed waypoints in a zig-zag pattern across the experiment site. During the flight, the UAV maintains a fixed height, and we repeat the experiments at different UAV heights, specifically 30 m, 50 m, 70 m, 90 m, and 110 m. To collect LTE I/Q samples from the BS tower and track the real-time location of the UAV, we equip the UAV with a portable node. The portable node consists of an SDR receiver and a GPS receiver. Regarding the configuration of the LTE evolved NodeB (eNB) at the BS tower, we set the tower height to 10~m, transmit power to 10~dBm, carrier frequency to 3.51~GHz, and bandwidth to 1.4~MHz. Both the BS tower and the UAV are equipped with USRP B205mini from National Instruments (NI). During the experiments, the UAV captures 20~ms segments of LTE signal out of every 100~ms. The snapshot of the experiment conducted at the AERPAW LWRFL is shown in Fig.~\ref{fig:experiment_site}. Fig.~\ref{fig:drone} shows the photo of the UAV during the experiment, where the downward RX antenna is installed at the bottom of the UAV, ensuring that the signal blockage from the airframe is minimized.

\begin{figure}[t!]
	\centering
	\includegraphics[width=0.48\textwidth]{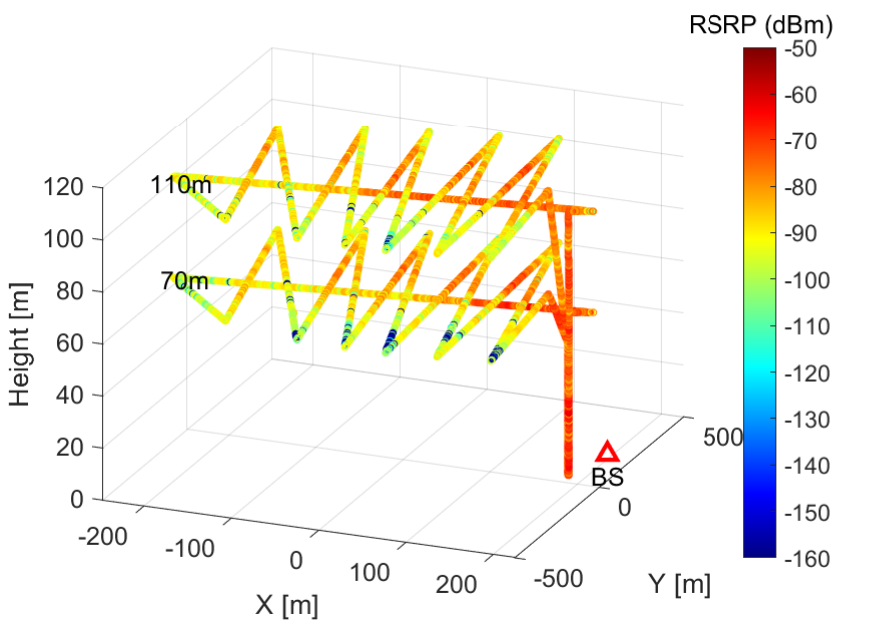}
	\caption{The trajectory of the UAV at two different altitudes. The RSRP is shown in different colors over the trajectory as indicated on the color bar. Collected LTE I/Q samples and GPS logs are post-processed to obtain the RSRP values. 
 %Multiple flights with different heights (70~m, 110~m) are overlapped.
 }\label{fig:trajectory}
\end{figure}

\subsection{LTE I/Q Samples Post-processing}\label{sec:post-process}
After we finish the experiments, we post-process the collected LTE I/Q samples using MATLAB's LTE Toolbox. The cell search and synchronization procedures are carried out by detecting the primary synchronization signal (PSS) and the secondary synchronization signal (SSS). Then, cell-specific reference signals (CRSs) are extracted by using physical cell identities (PCIs). In the end, RSRP is calculated using the CRSs, which we use as the main indicator for the received signal strength. In addition to the data we exctract from the I/Q samples, we also use the GPS log information from the UAV to extract the time and location information for each I/Q sample. 

Fig.~\ref{fig:spectro} shows the spectrogram for a $16$~ms duration of the collected IQ samples with $2$~MHz sampling frequency. Note that there are no users in active communications with the eNB, and hence, no data is being transmitted in the downlink. In other words, the spectrum is occupied only by evolved node B (eNB)'s physical broadcast channel (PBCH), the CRSs, the PSS, and SSS transmissions in the downlink, and the rest of the spectrum consists of noise plus interference. We observe that noise plus interference power from the UAV's self-emissions~\cite{pienaar2016rf} is sufficiently low to decode the received LTE signal from the ground BS. Moreover, no other interference sources at the frequency band we utilize for the experiments are detected during the flights.

Fig.~\ref{fig:trajectory} shows the UAV trajectory along with the RSRPs from the measurement dataset plotted with different colors through the UAV's trajectory. Note that other than having different altitudes, the UAV trajectories are otherwise identical.  
%The color of the trajectory indicates the received signal strength (RSRP). 
It is clearly observed that the UAV flies up to a certain height and sweeps the experiment area by the zig-zag pattern while holding its height. After reaching the final waypoint, the UAV directly returns back to the landing point. A comprehensive discussion on how the data is post-processed and additional related results can be found in~\cite{maeng2022aeriq,maeng2023lte}.

\section{Air-to-ground Propagation Model}\label{Sec:A2G_propagation_model}
In this section, we consider path loss models and 3D antenna patterns to model and analyze the received signal strength. The received signal strength can be expressed as
\begin{align}\label{eq:received}
    r&=\mathsf{P}_{\rm Tx}-\mathsf{PL}+s,
\end{align}
where $\mathsf{P}_{\rm Tx}$, $\mathsf{PL}$, and $s$ denote the transmit power, the path loss, and the shadowing component, respectively.

\begin{figure}[t]
	\centering
	\includegraphics[width=0.48\textwidth]{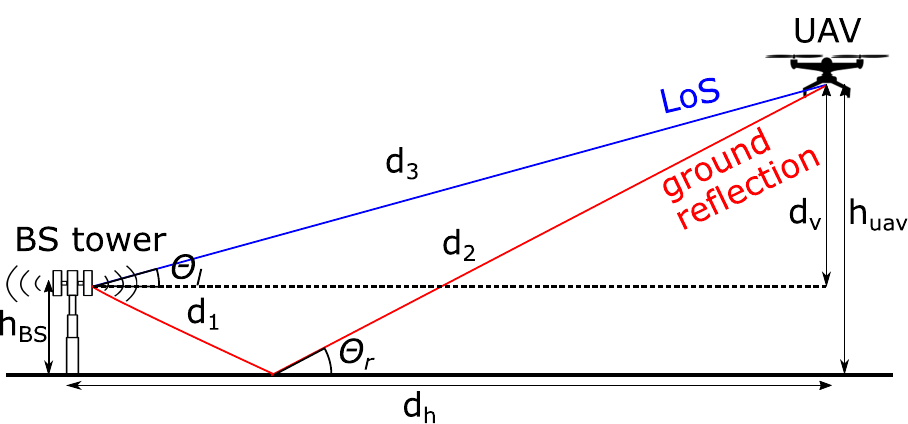}
	\caption{The illustration of the two-ray ground reflection model.}\label{fig:two_ray_illu}
\end{figure}

\subsection{Path Loss Model}
To characterize the air-to-ground channel in the rural environment, we adopt the two-ray path loss model between a BS tower and a UAV. This model considers a
line-of-sight (LoS) path as well as a strong ground reflection path, both contributing to the received signal. The path loss characterized by the two-ray ground reflection model can be expressed as~\cite[Chapter 2]{jakes1994microwave}
\begin{align}\label{eq:PL_two}
    &\mathsf{PL}_{\rm 2R}=\left(\frac{\lambda}{4\pi}\right)^2\times\left|\frac{\sqrt{\mathsf{G}_{\rm Tx}(\phi_l,\theta_l)\mathsf{G}_{\rm Rx}(\phi_l,\theta_l)}}{d_{\rm 3}}\right.\nonumber\\
    &\left.+\frac{\Gamma(\theta_r)\sqrt{\mathsf{G}_{\rm Tx}(\phi_r,\theta_r)\mathsf{G}_{\rm Rx}(\phi_r,\theta_r)}e^{-j\Delta\tau}}{d_1+d_2}\right|^2,
\end{align}
where $\mathsf{G}_{\rm Tx}(\phi,\theta)$, $\mathsf{G}_{\rm Rx}(\phi,\theta)$ denote the antenna gain of a transmitter and a receiver from 3D antenna radiation patterns depending on azimuth ($\phi$) and elevation ($\theta$) angles, $\lambda$, $\theta_r=\tan^{-1}\left(\frac{h_{\rm BS}+h_{\rm uav}}{d_{\rm h}}\right)$ indicate wave-length and the ground reflection angle, and $\Delta\tau=\frac{2\pi(d_1+d_2-d_{3})}{\lambda}$ indicates the phase difference of two paths at the UAV. The distance and the angle parameters in the two-ray ground reflection model are described in Fig.~\ref{fig:two_ray_illu}. The ground reflection coefficient of the vertically polarized signal can be modeled as
\begin{align}
    \Gamma(\theta_r)&=\frac{\varepsilon_0\sin\theta_r-\sqrt{\varepsilon_0-\cos^2\theta_r}}{\varepsilon_0\sin\theta_r+\sqrt{\varepsilon_0-\cos^2\theta_r}},
\end{align}
where $\varepsilon_0$ is the relative permittivity of the ground, which can be varied depending on the ground condition. If we only consider the LoS path in the two-ray path loss, we can obtain the free-space path loss model, given as
\begin{align}\label{eq:PL_fs}
    &\mathsf{PL}_{\rm FS}=\left(\frac{\lambda}{4\pi}\right)^2\left|\frac{\sqrt{\mathsf{G}_{\rm Tx}(\phi_l,\theta_l)\mathsf{G}_{\rm Rx}(\phi_l,\theta_l)}}{d_{3}}\right|^2.
\end{align}

\begin{figure}[t]
	\centering
	\includegraphics[width=0.4\textwidth]{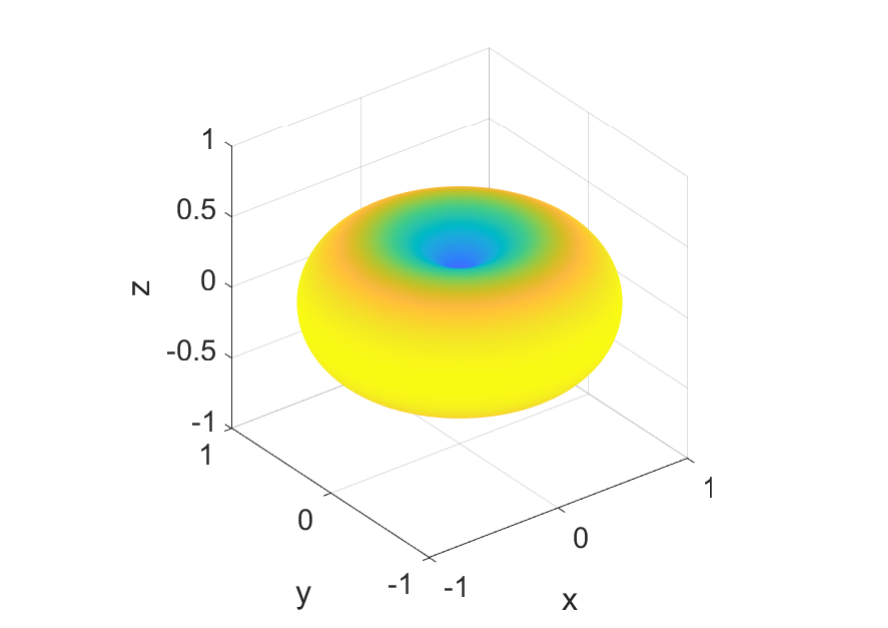}
	\caption{The 3D dipole antenna pattern of \eqref{eq:dipole_formula} used as the second antenna pattern option.}\label{fig:dipole_pat}
\end{figure}

\begin{figure*}[t!]
	\centering
	\subfloat[Anechoic chamber setup for Tx antenna pattern measurement. The center point of the chamber is adjusted to the tip of the antenna by the crossed laser lines.]{\includegraphics[width=0.33\textwidth]{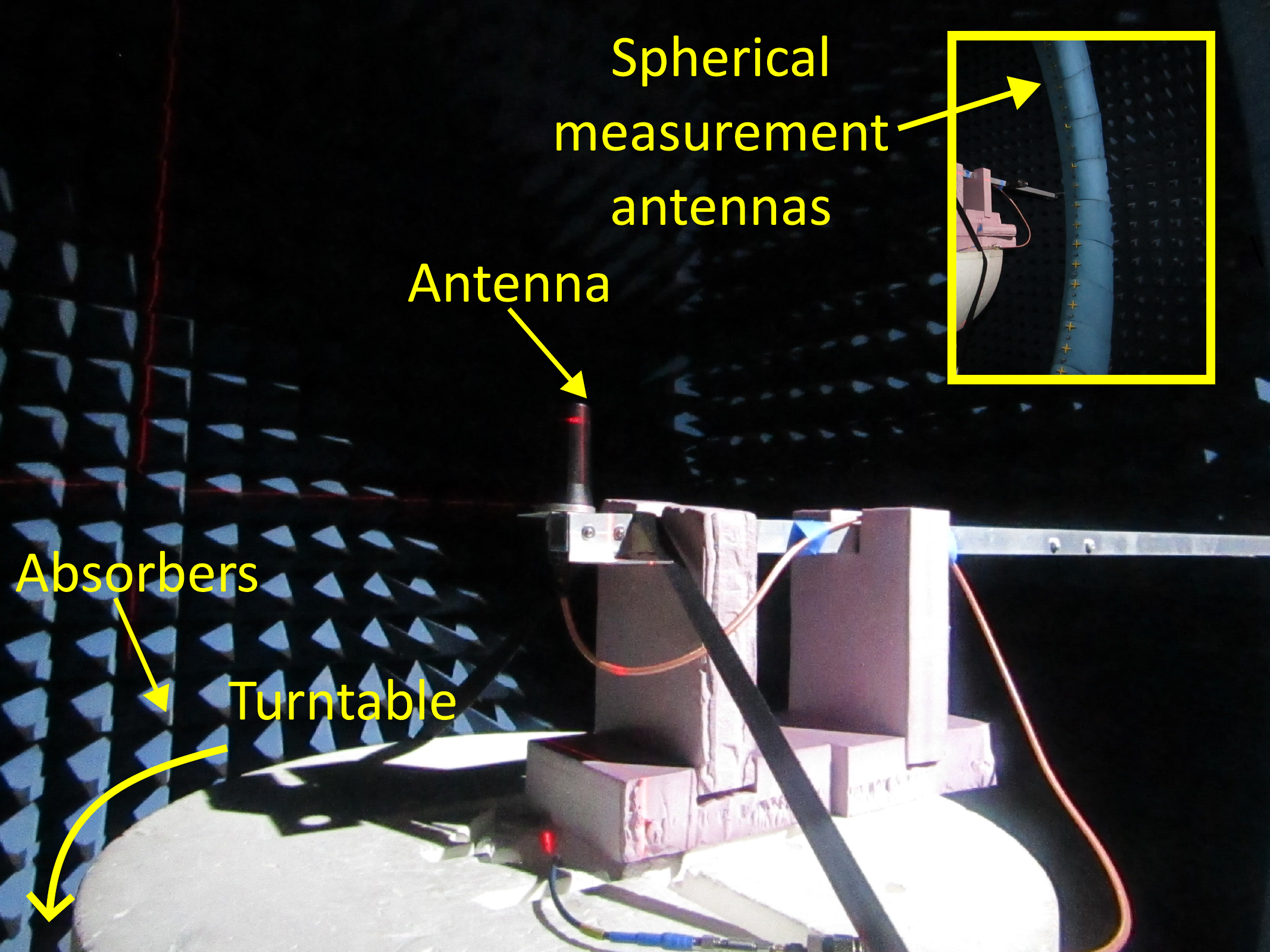}\label{fig:chamber}}~
        \subfloat[Measured Tx antenna pattern in Cartesian coordinates by the linear scale antenna gain.]{\includegraphics[width=0.33\textwidth]{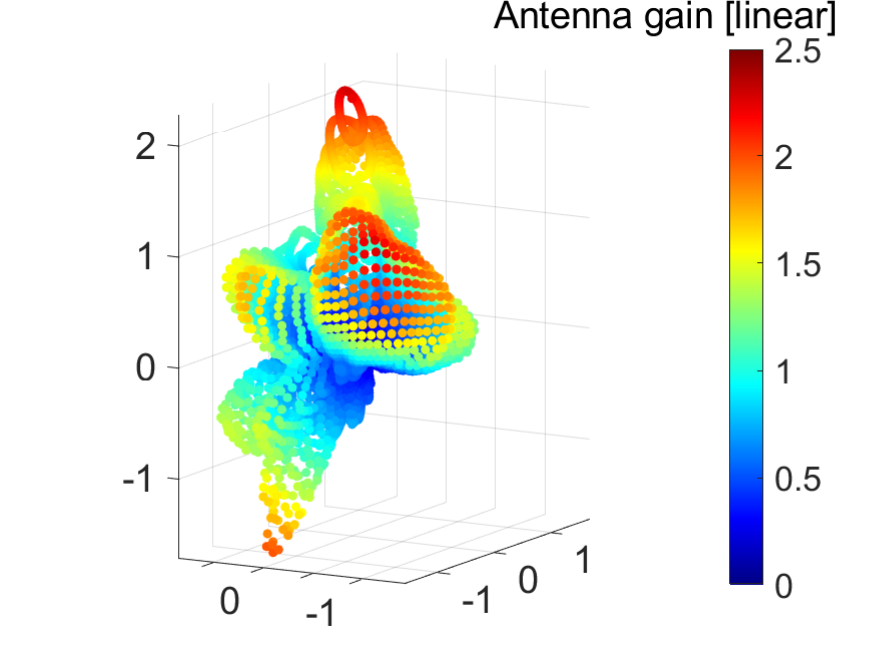}\label{fig:Tx_ant_pat}}~
        \subfloat[Tx antenna pattern in 2D angles domain by dB scale antenna gain.]{\includegraphics[width=0.33\textwidth]{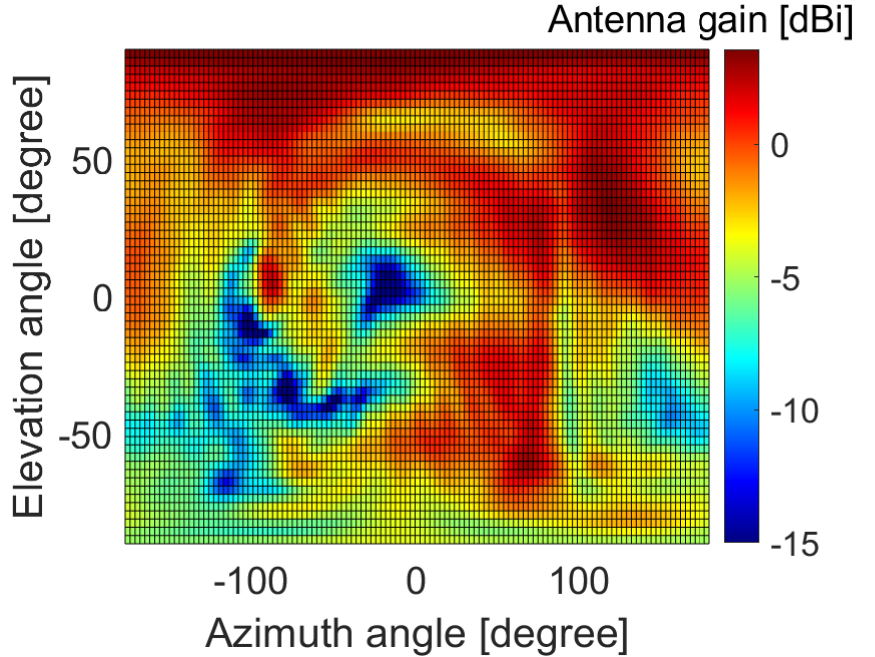}\label{fig:Tx_ant_pat_2}}

        \subfloat[Measured Tx antenna radiation pattern in elevation angle domain.]{\includegraphics[width=0.33\textwidth]{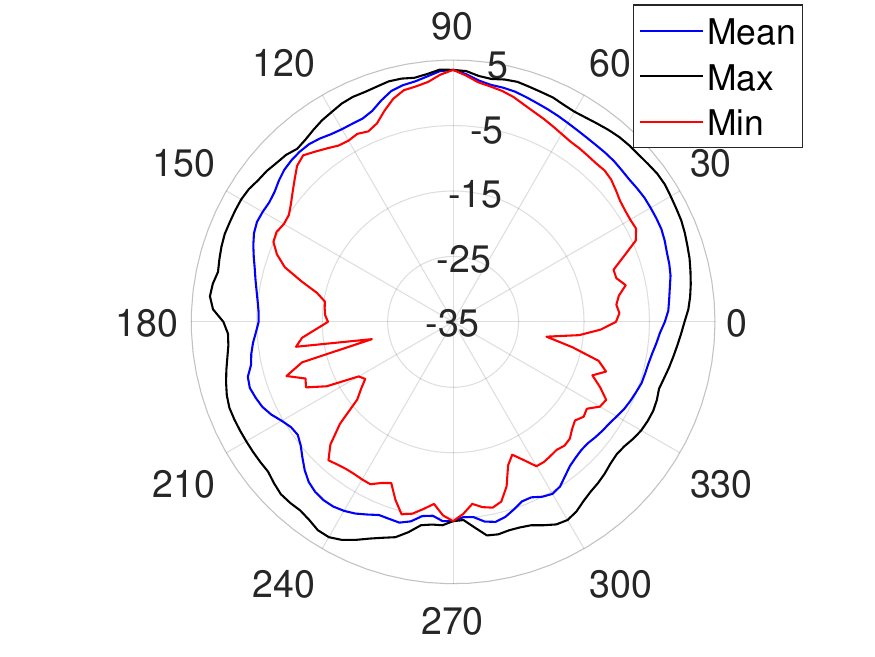}\label{fig:Tx_ant_pat_elev}}~
        \subfloat[Rx antenna pattern for the elevation angle domain by the dB scale in the specification sheet.]{\includegraphics[width=0.33\textwidth]{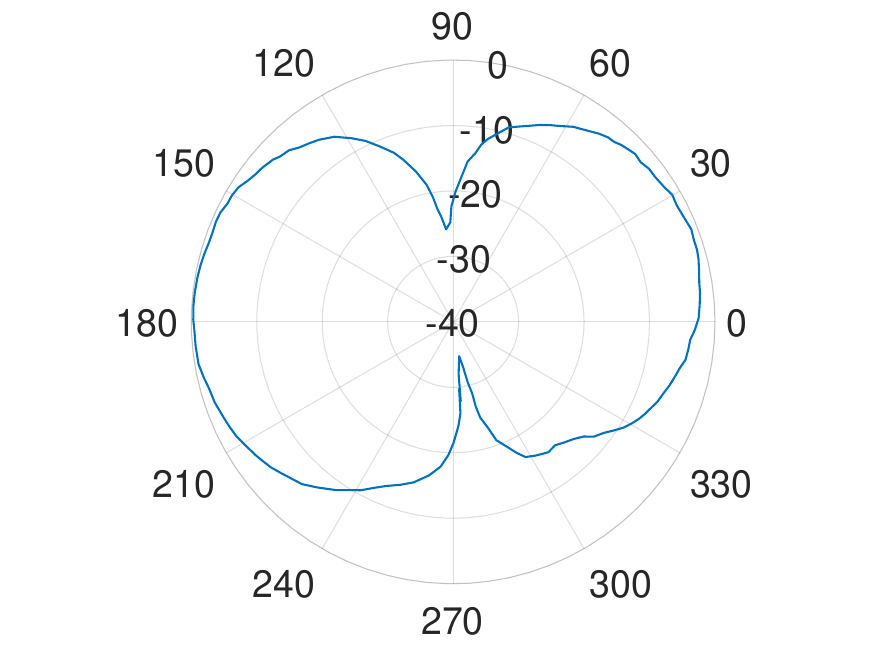}\label{fig:Rx_ant_pat}}~
        \subfloat[Measured Rx antenna radiation pattern by the anechoic chamber.]{\includegraphics[width=0.33\textwidth]{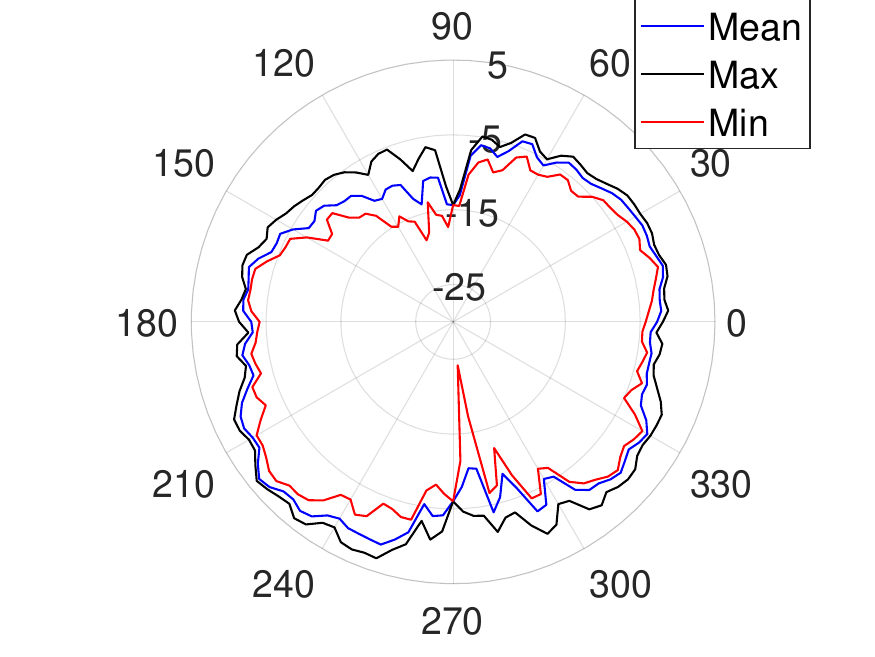}\label{fig:Rx_ant_pat_mea}}
	\caption{The photo of the anechoic chamber setup for the 3D antenna pattern measurement, the Tx, Rx antenna patterns we used for the analysis, and the antenna installment of the UAV.}\label{fig:ant_pat}
\end{figure*}

\begin{figure*}[t!]
	\centering
	\subfloat[Measured antenna pattern, 30~m.]{\includegraphics[width=0.33\textwidth]{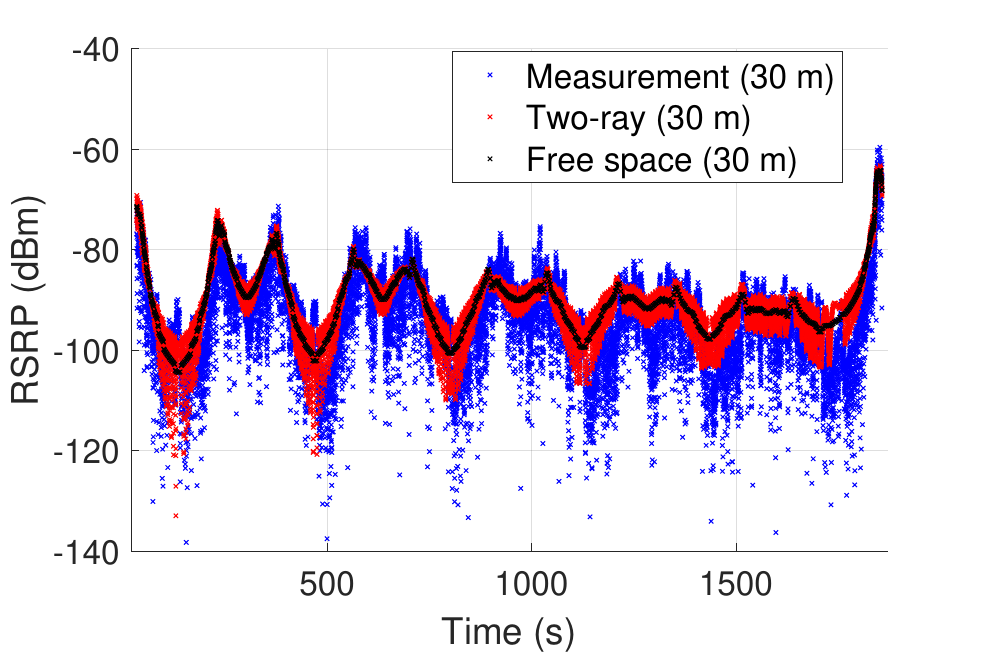}\label{fig:RSRP_t_30_1}}~
	\subfloat[Measured antenna pattern, 70~m.]{\includegraphics[width=0.33\textwidth]{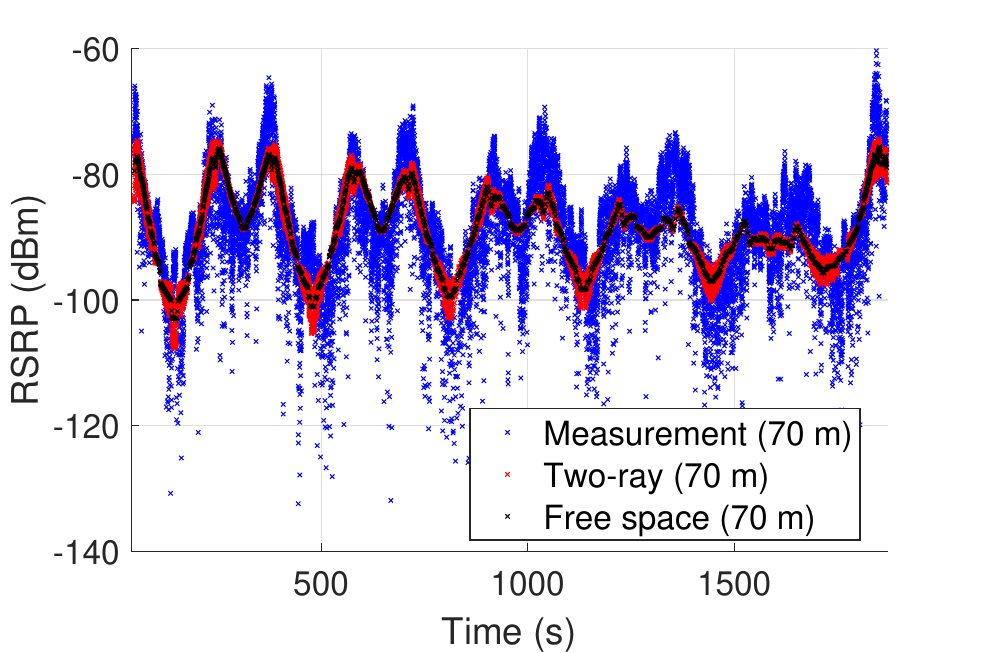}\label{fig:RSRP_t_70_1}}~
        \subfloat[Measured antenna pattern, 110~m.]{\includegraphics[width=0.33\textwidth]{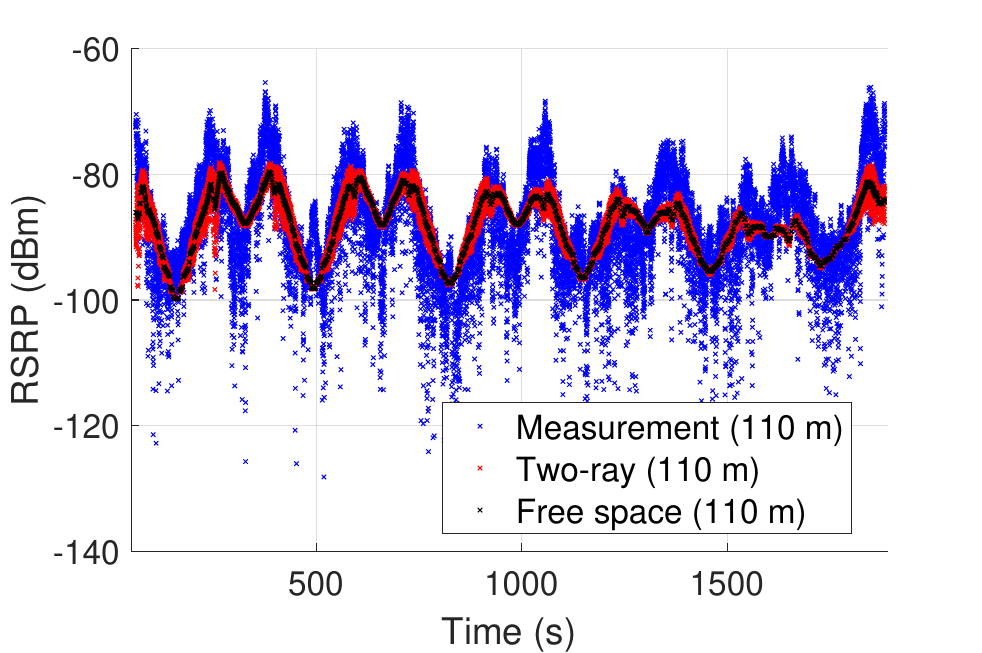}\label{fig:RSRP_t_110_1}}
        \vspace{-0.02in}
        \subfloat[Dipole antenna pattern, 30~m.]{\includegraphics[width=0.33\textwidth]{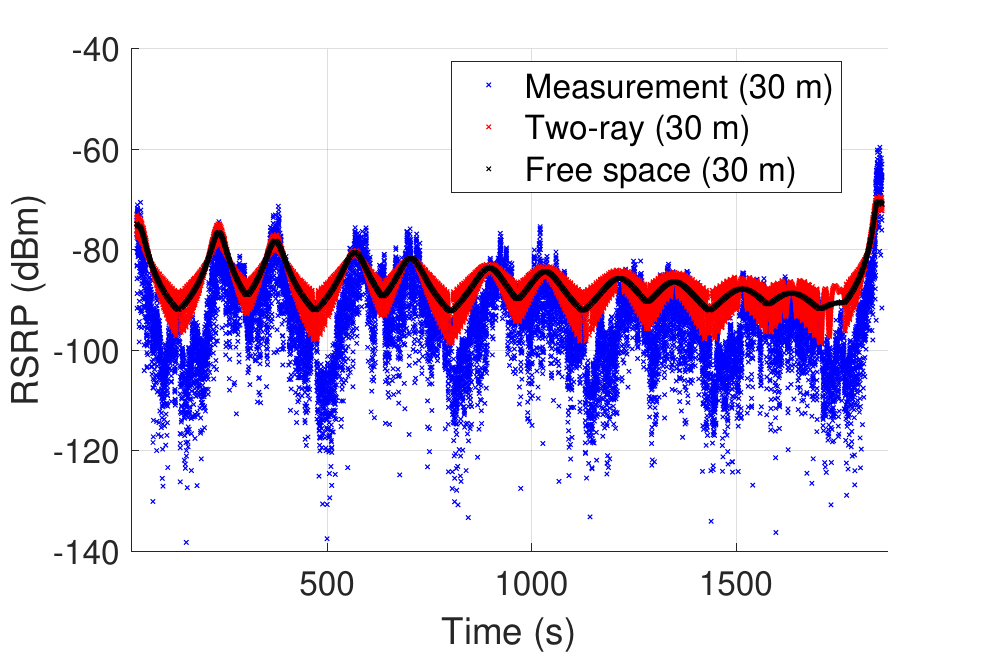}\label{fig:RSRP_t_30_2}}~
	\subfloat[Dipole antenna pattern, 70~m.]{\includegraphics[width=0.33\textwidth]{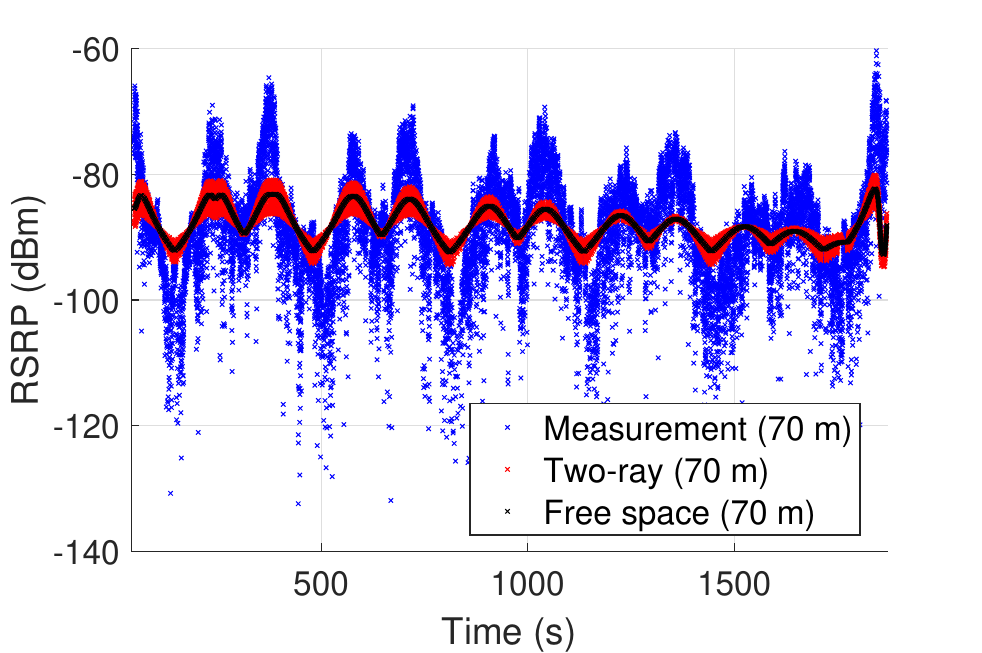}\label{fig:RSRP_t_70_2}}~
        \subfloat[Dipole antenna pattern, 110~m.]{\includegraphics[width=0.33\textwidth]{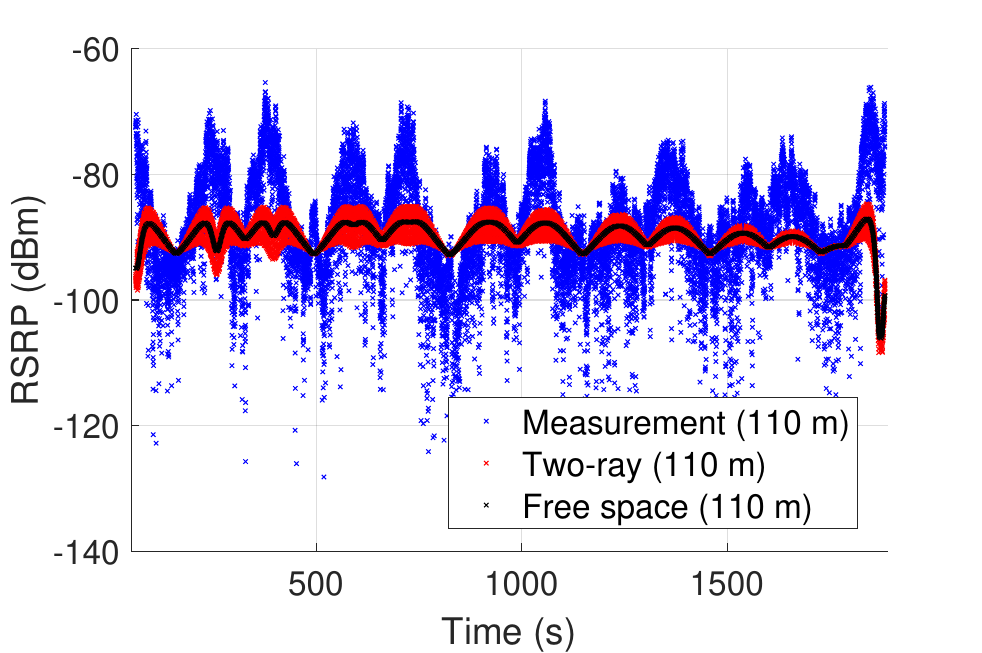}\label{fig:RSRP_t_110_2}}
        \vspace{-0.02in}
        \subfloat[isotropic antenna pattern, 30~m.]{\includegraphics[width=0.33\textwidth]{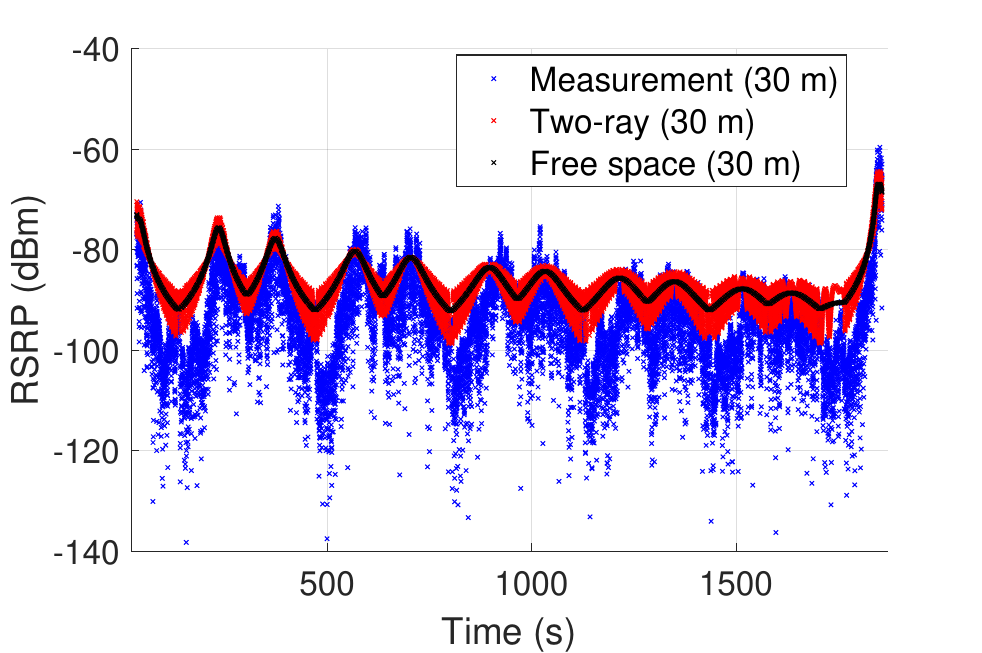}\label{fig:RSRP_t_30_3}}~
	\subfloat[isotropic antenna pattern, 70~m.]{\includegraphics[width=0.33\textwidth]{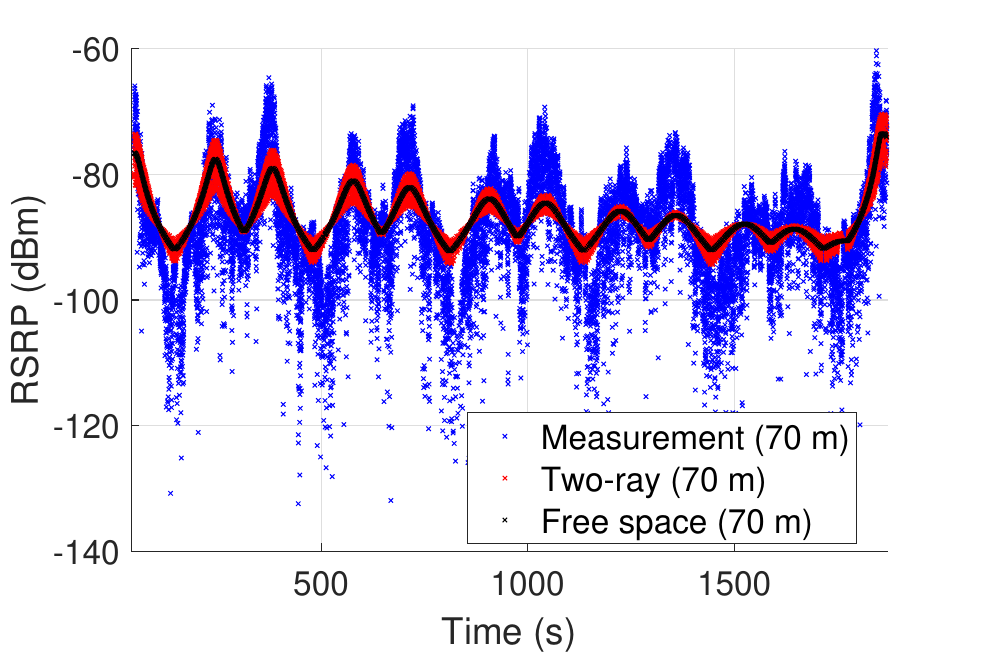}\label{fig:RSRP_t_70_3}}~
        \subfloat[isotropic antenna pattern, 110~m.]{\includegraphics[width=0.33\textwidth]{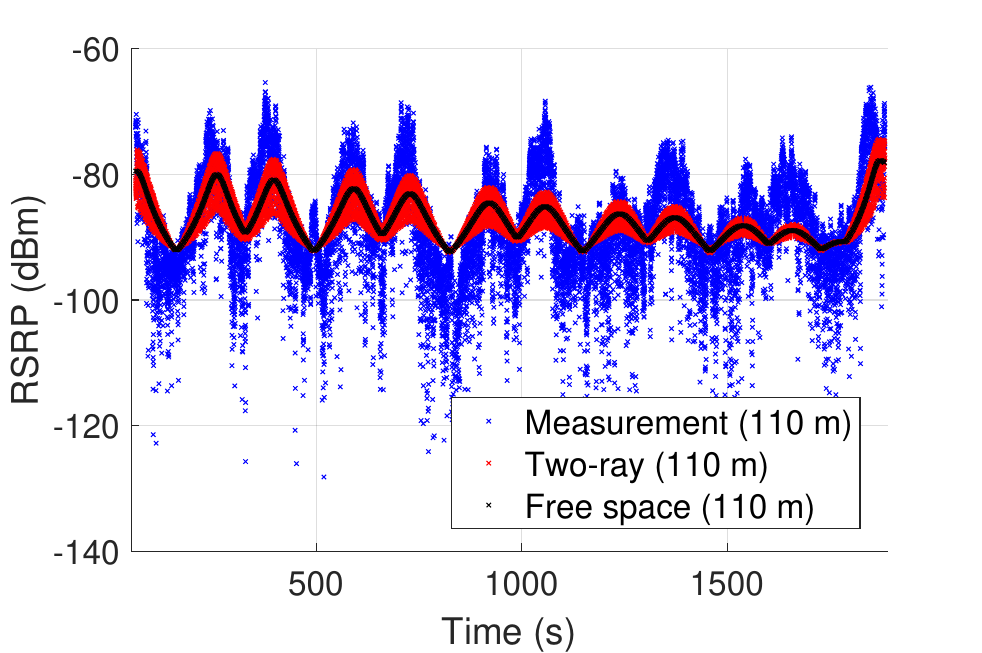}\label{fig:RSRP_t_110_3}}
	\caption{Time domain comparison between measured RSRP and analytical RSRP with different path loss models and antenna patterns.}\label{fig:RSRP_t}
\end{figure*}

\begin{figure*}[t!]
	\centering
	\subfloat[Measured antenna pattern, 30~m.]{\includegraphics[width=0.33\textwidth]{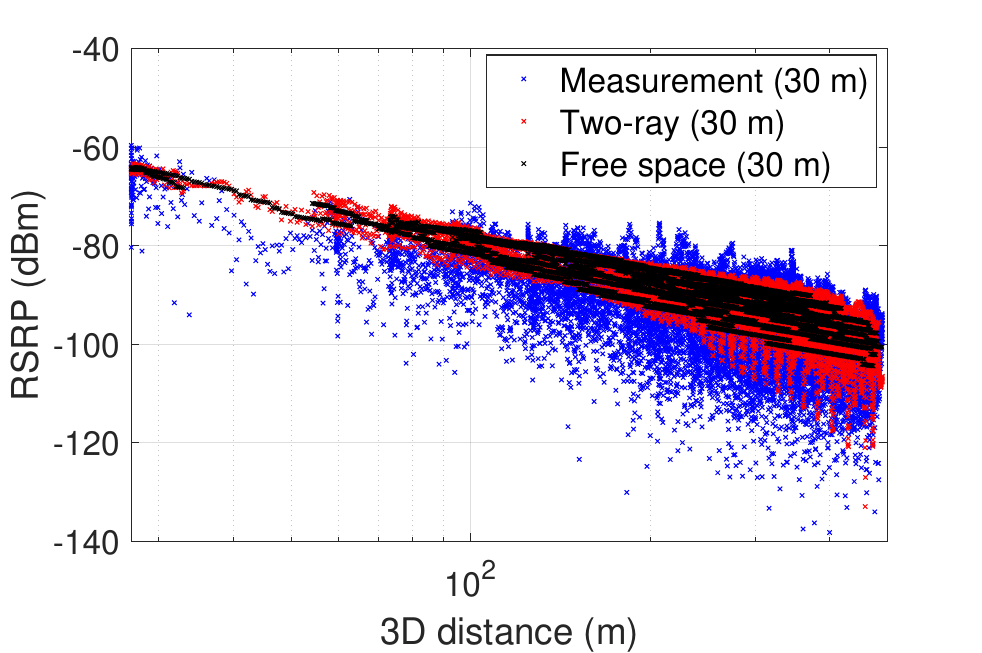}\label{fig:}}~
	\subfloat[Measured antenna pattern, 70~m.]{\includegraphics[width=0.33\textwidth]{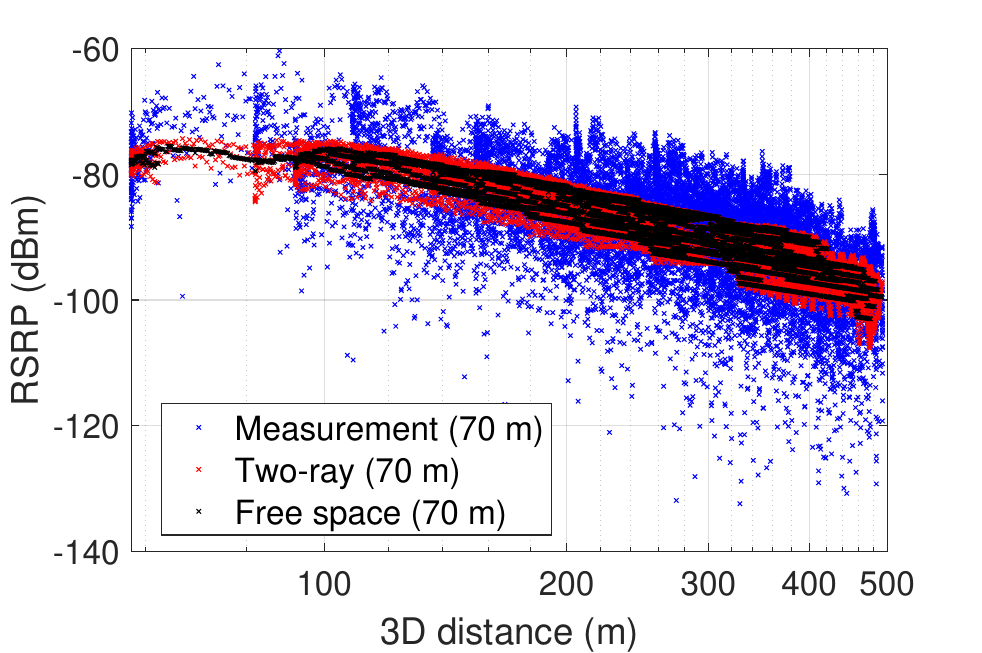}\label{fig:}}~
        \subfloat[Measured antenna pattern, 110~m.]{\includegraphics[width=0.33\textwidth]{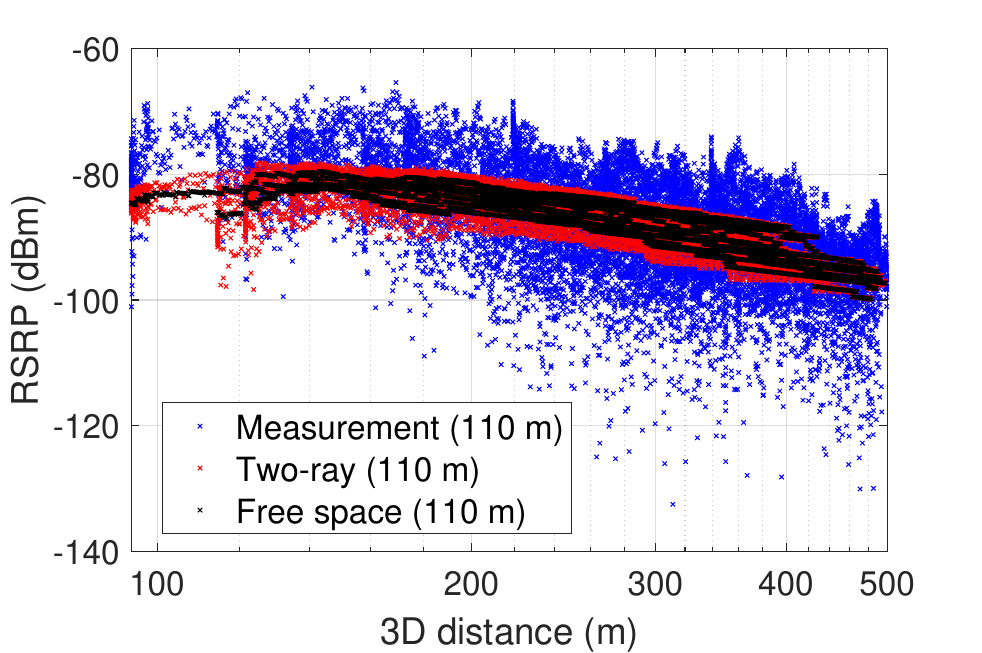}}
        \vspace{-0.02in}
        \subfloat[Dipole antenna pattern, 30~m.]{\includegraphics[width=0.33\textwidth]{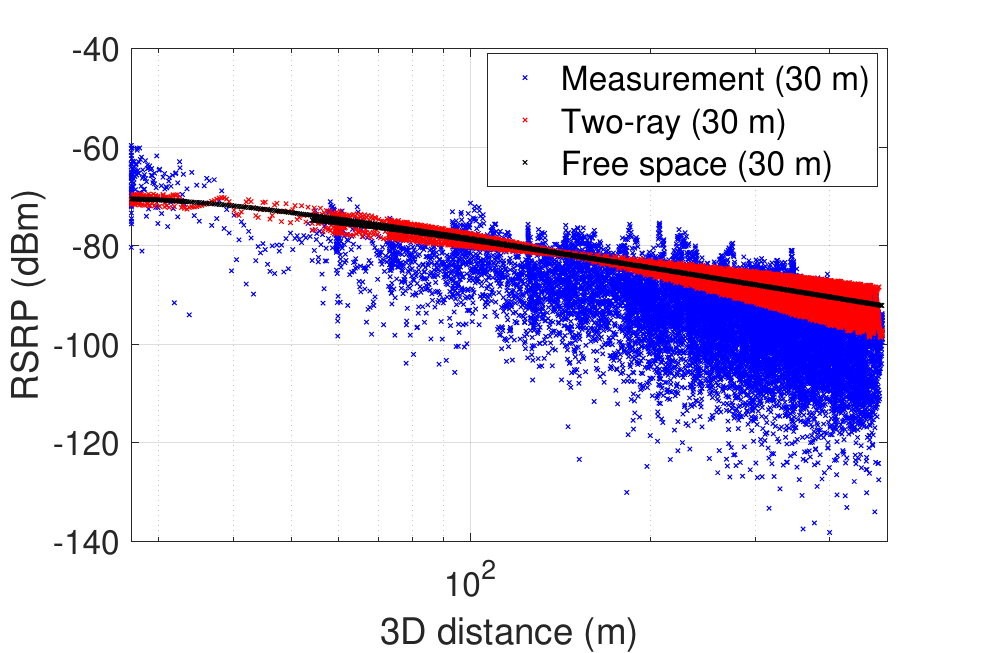}}~
	\subfloat[Dipole antenna pattern, 70~m.]{\includegraphics[width=0.33\textwidth]{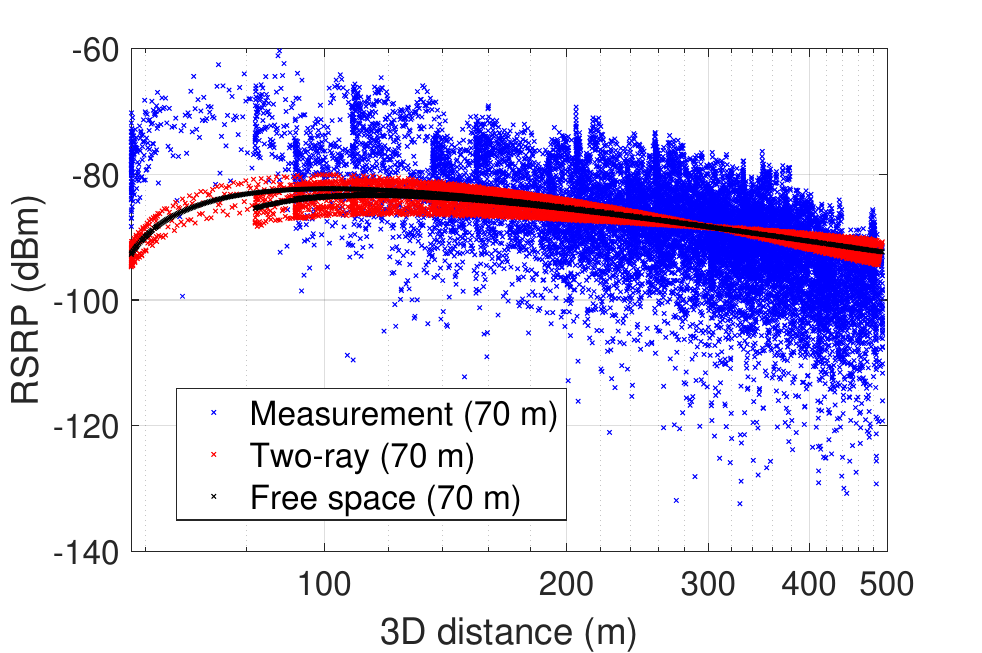}}~
        \subfloat[Dipole antenna pattern, 110~m.]{\includegraphics[width=0.33\textwidth]{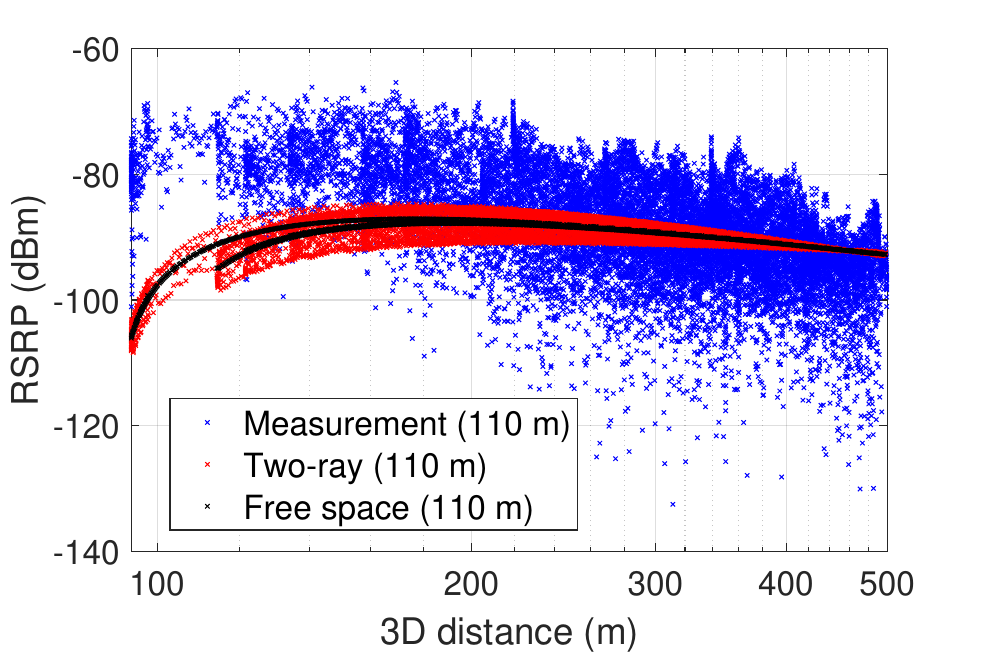}}
        \vspace{-0.02in}
        \subfloat[isotropic antenna pattern, 30~m.]{\includegraphics[width=0.33\textwidth]{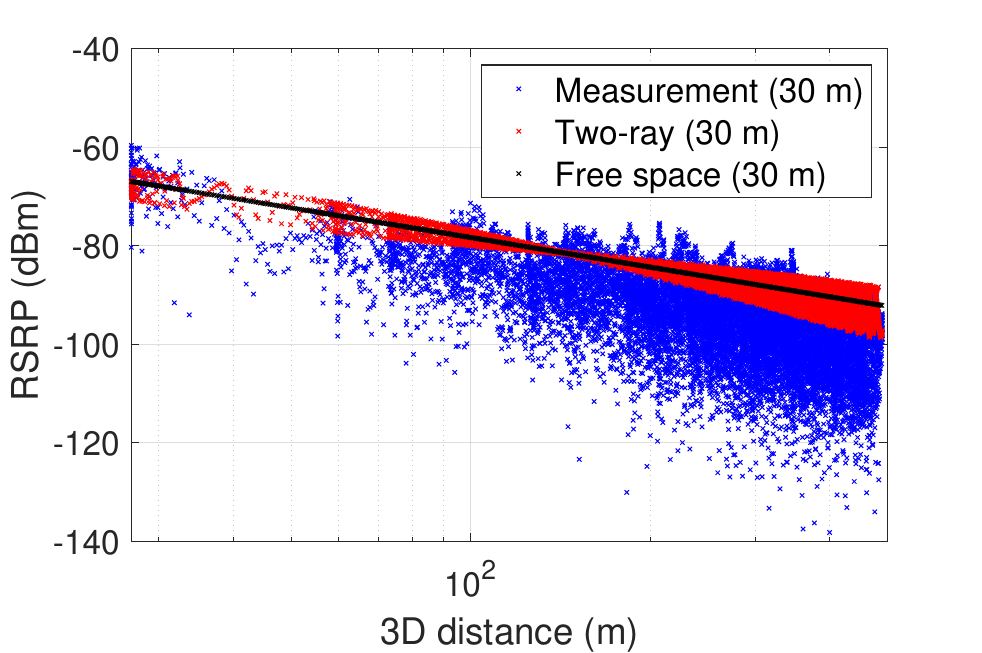}}~
	\subfloat[isotropic antenna pattern, 70~m.]{\includegraphics[width=0.33\textwidth]{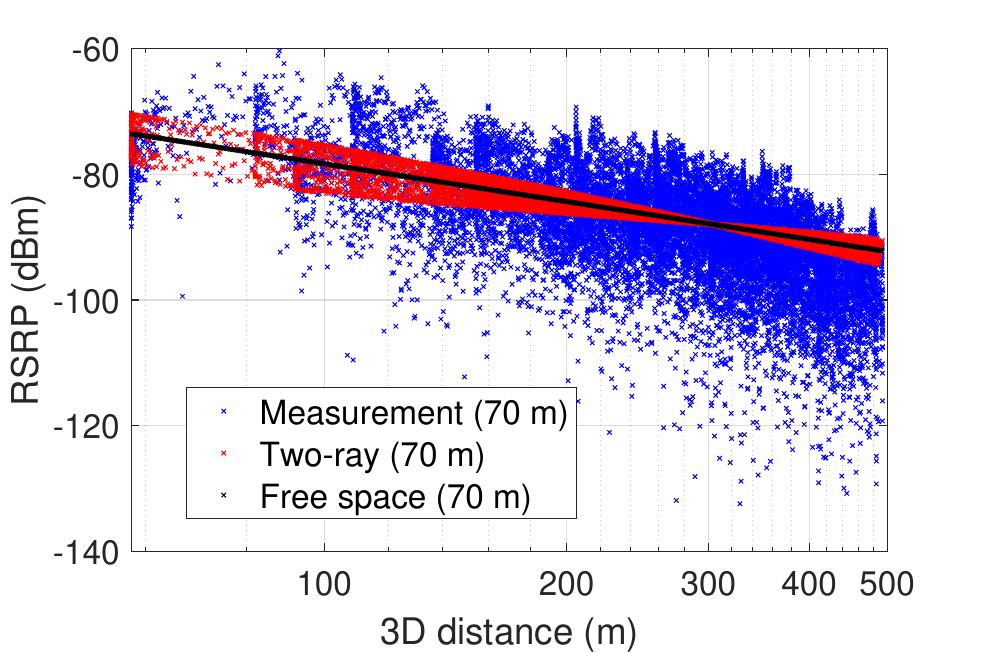}}~
        \subfloat[isotropic antenna pattern, 110~m.]{\includegraphics[width=0.33\textwidth]{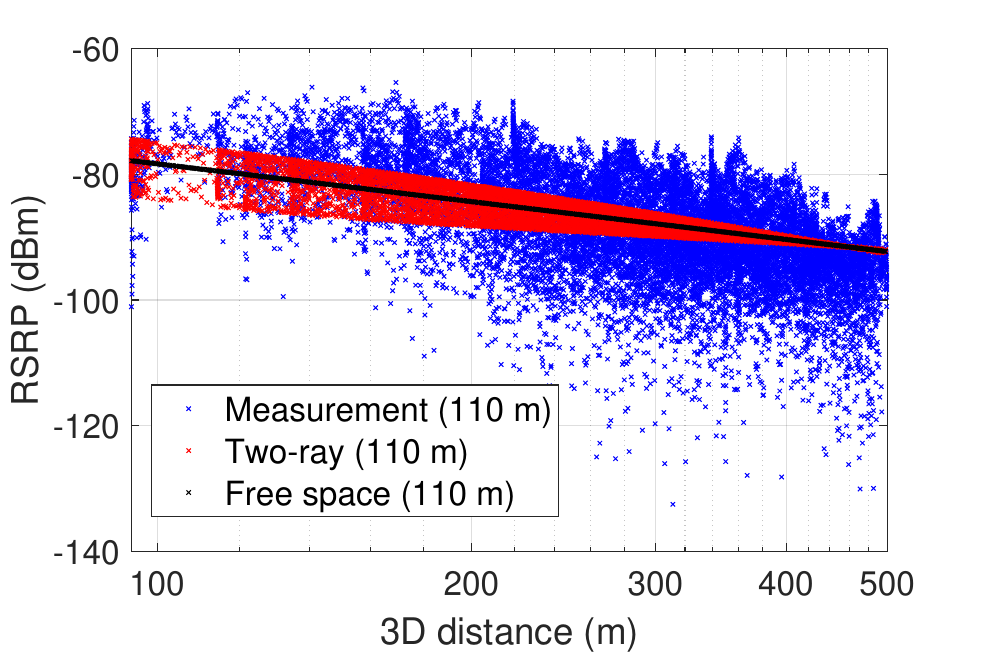}}
	\caption{Distance domain comparison between measured RSRP and analytical RSRP with different path loss models and antenna patterns.}\label{fig:RSRP_d}
\end{figure*}

\begin{figure*}[t!]
	\centering
	\subfloat[Measured antenna pattern, 30~m.]{\includegraphics[width=0.31\textwidth]{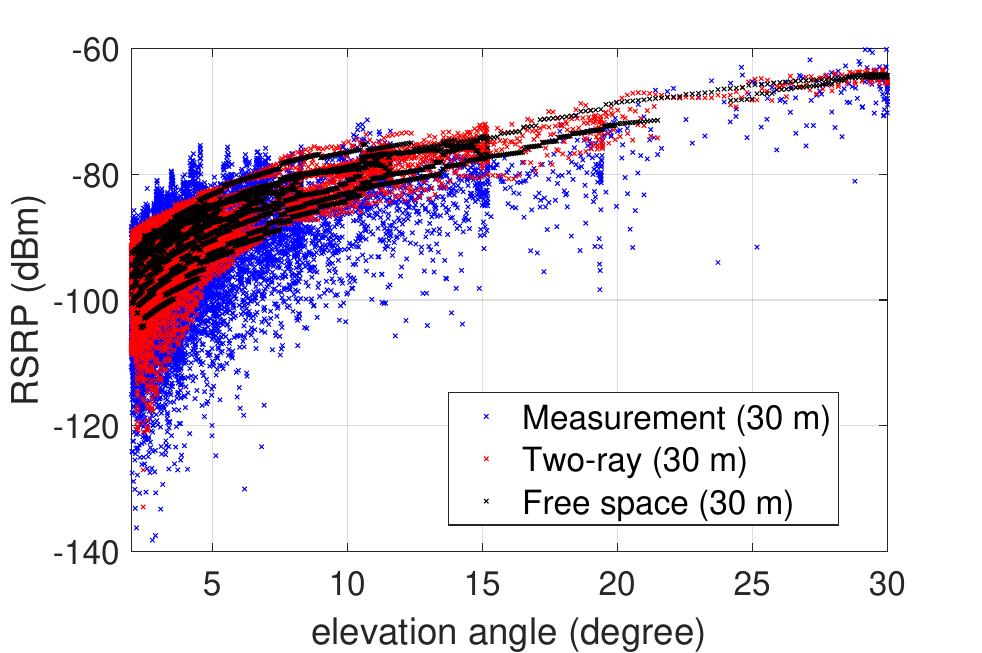}\label{fig:}}~
	\subfloat[Measured antenna pattern, 70~m.]{\includegraphics[width=0.31\textwidth]{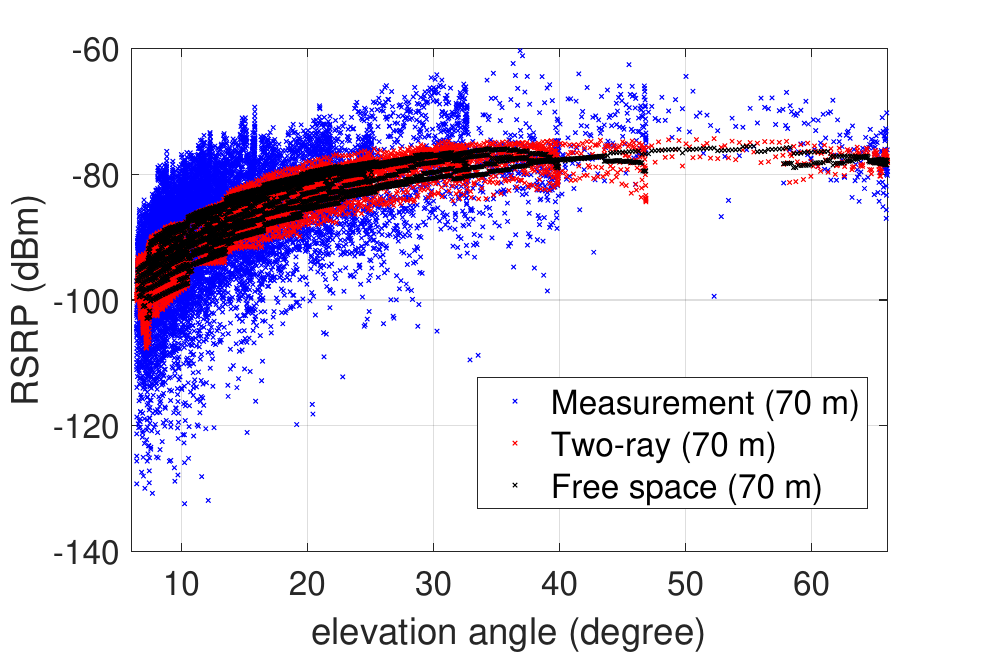}\label{fig:}}~
        \subfloat[Measured antenna pattern, 110~m.]{\includegraphics[width=0.31\textwidth]{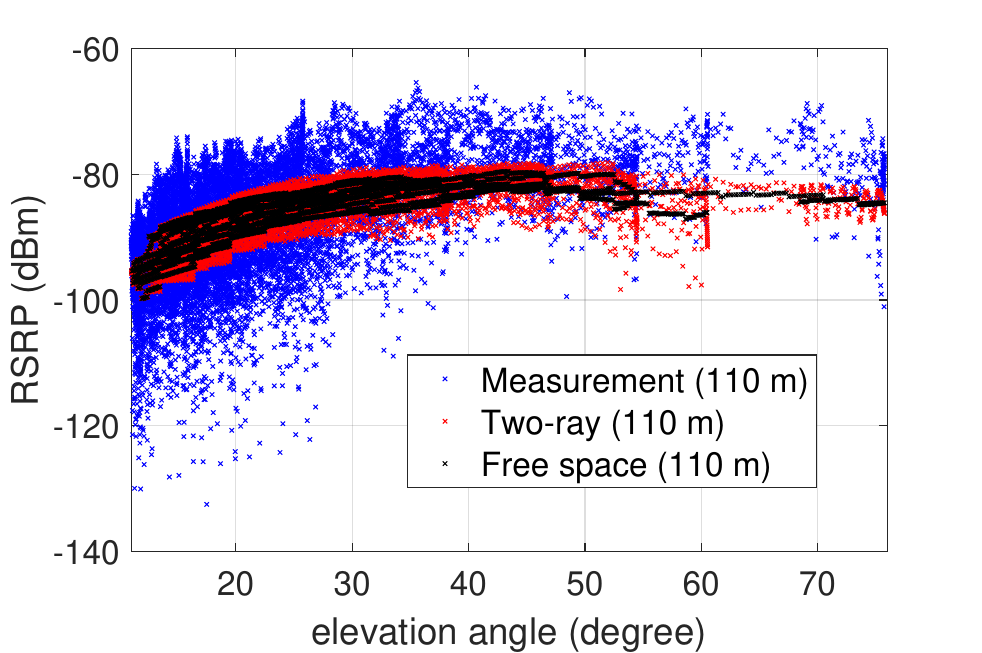}}
        \vspace{-0.02in}
        \subfloat[Dipole antenna pattern, 30~m.]{\includegraphics[width=0.31\textwidth]{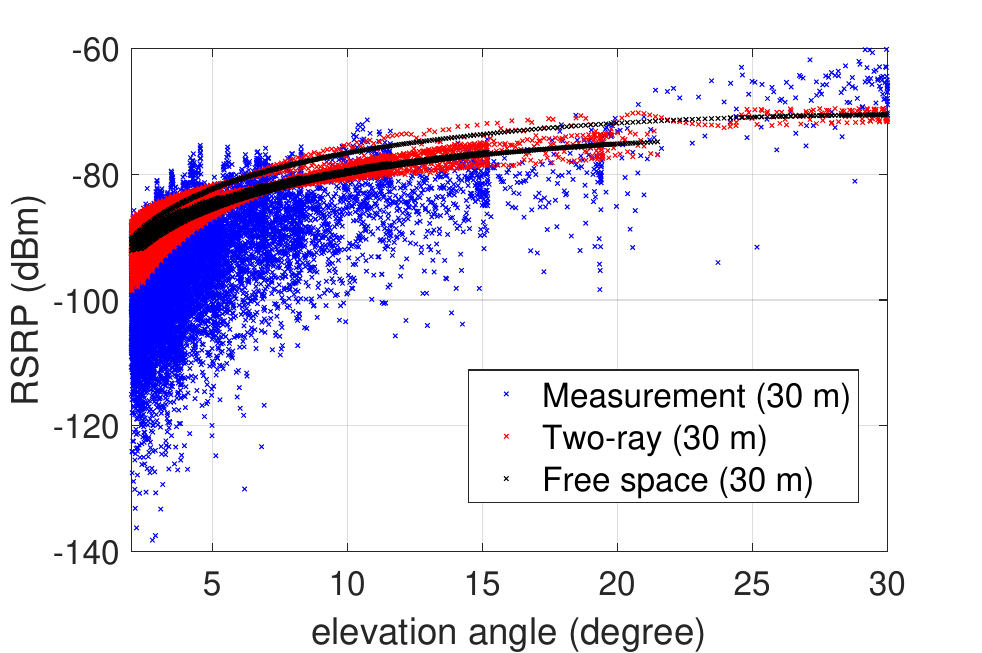}}~
	\subfloat[Dipole antenna pattern, 70~m.]{\includegraphics[width=0.31\textwidth]{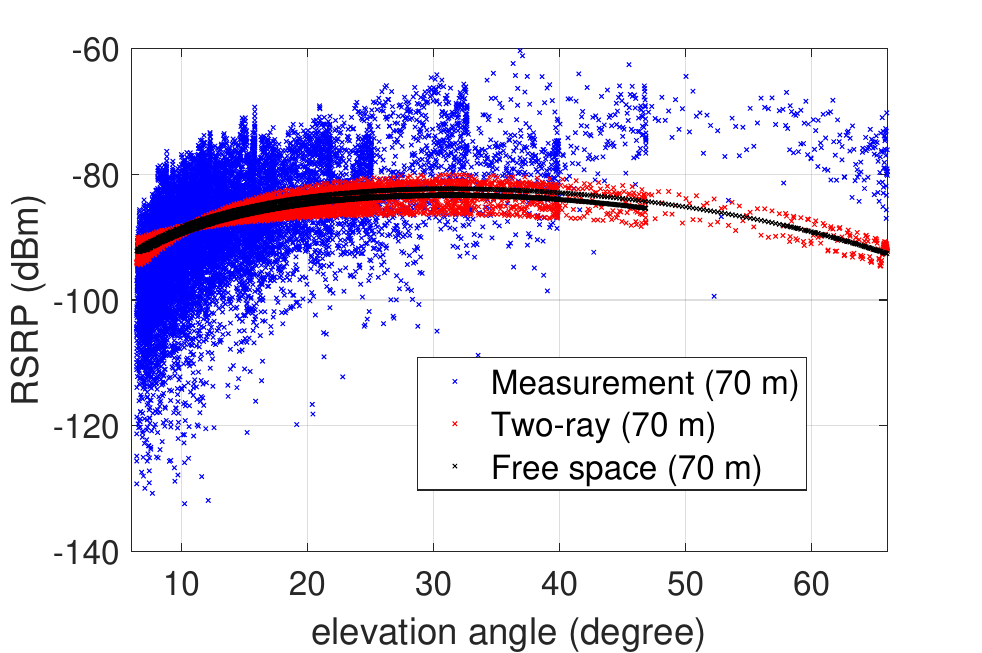}}~
        \subfloat[Dipole antenna pattern, 110~m.]{\includegraphics[width=0.31\textwidth]{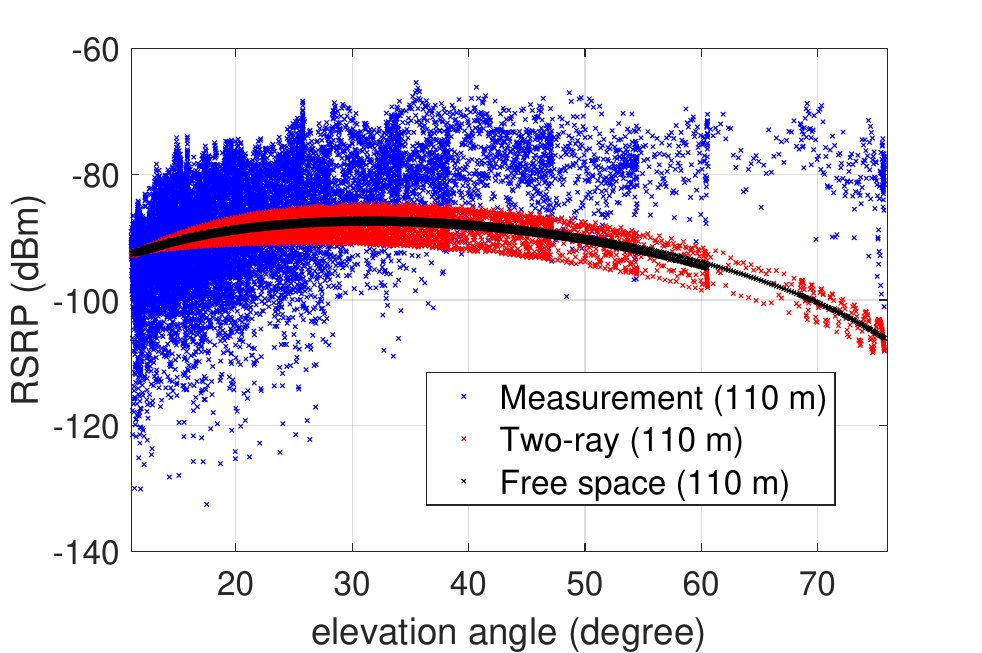}}
        \vspace{-0.02in}
        \subfloat[isotropic antenna pattern, 30~m.]{\includegraphics[width=0.31\textwidth]{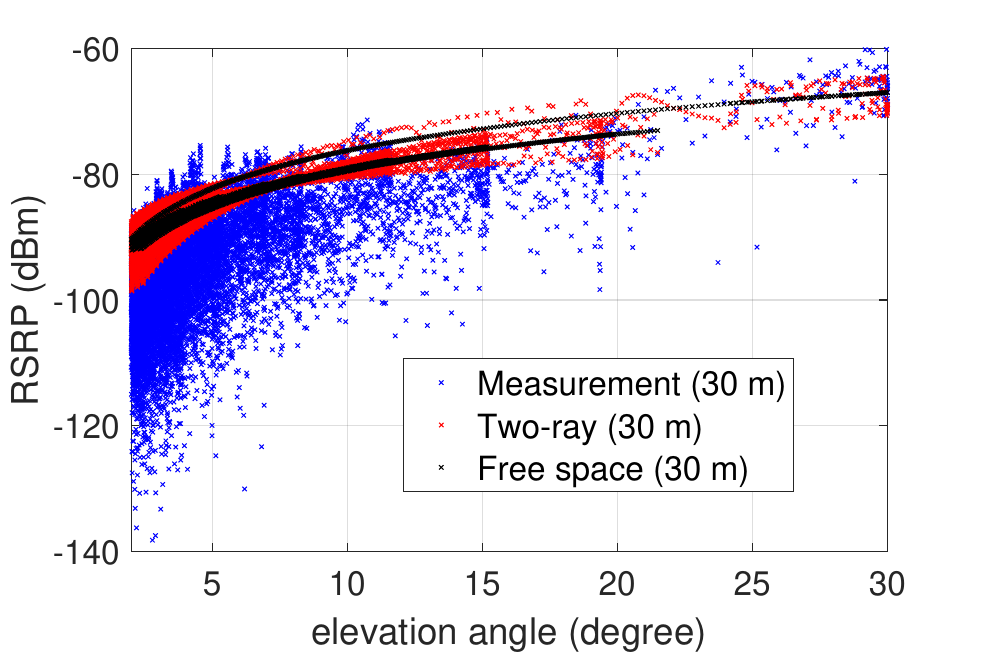}}~
	\subfloat[isotropic antenna pattern, 70~m.]{\includegraphics[width=0.31\textwidth]{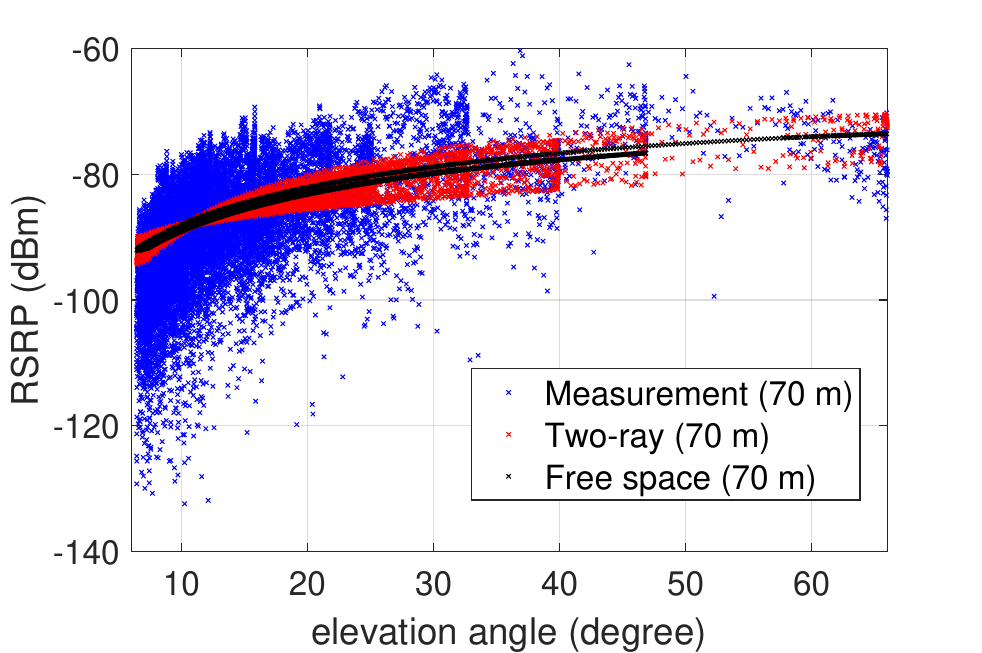}}~
        \subfloat[isotropic antenna pattern, 110~m.]{\includegraphics[width=0.31\textwidth]{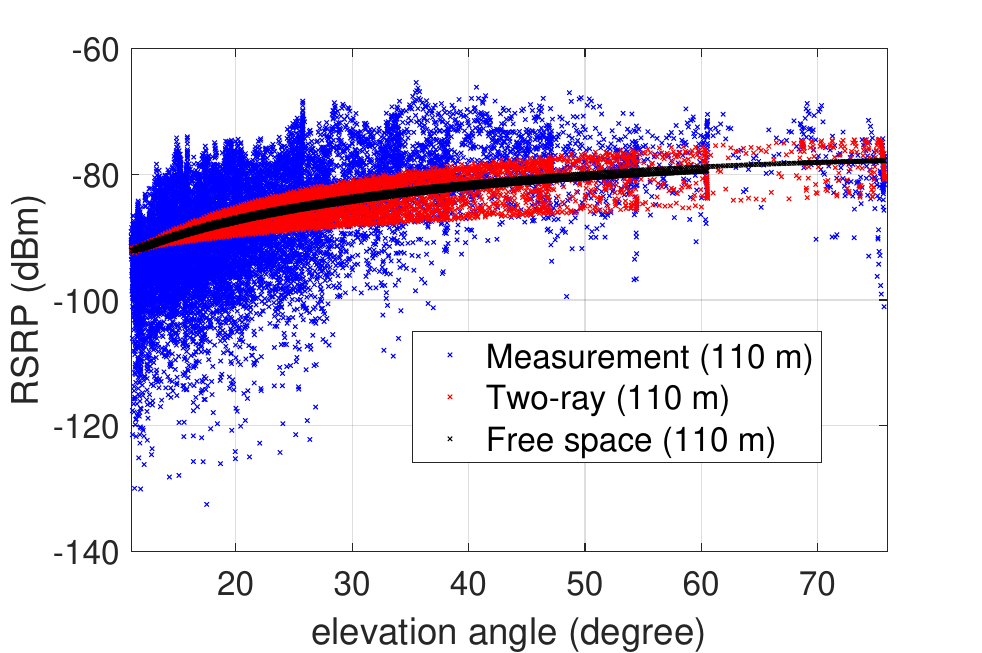}}
	\caption{Elevation angle domain comparison between measured RSRP and analytical RSRP with different path loss models and antenna patterns.}\label{fig:RSRP_e}
\end{figure*}

\begin{figure*}[t]
	\centering
	\subfloat[CDF of RSRP, 30~m.]{\includegraphics[width=0.33\textwidth]{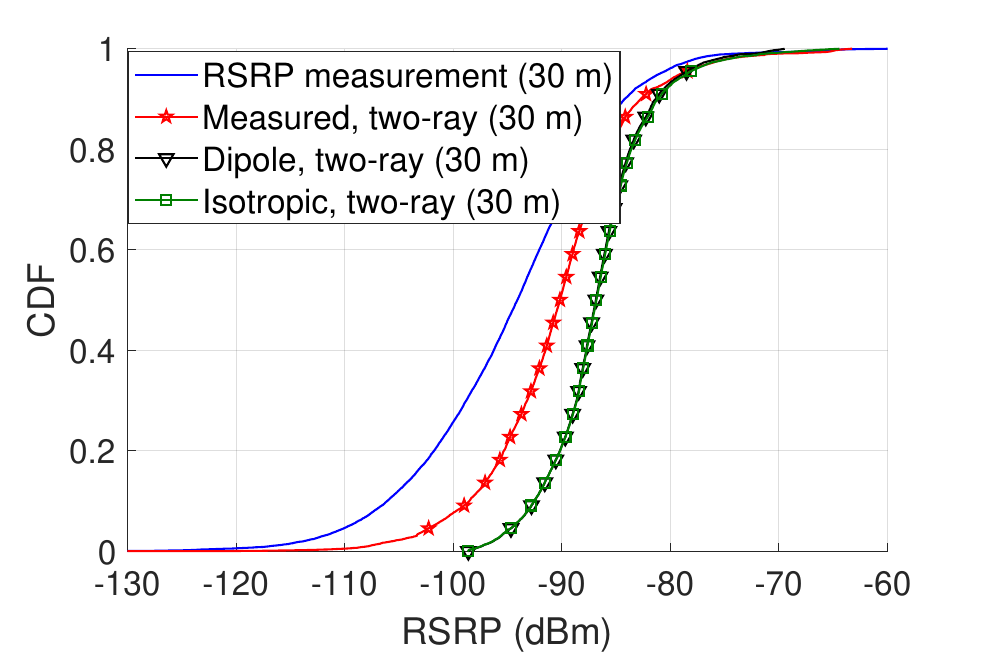}\label{fig:RSRP_cdf_30}}~
	\subfloat[CDF of RSRP, 70~m.]{\includegraphics[width=0.33\textwidth]{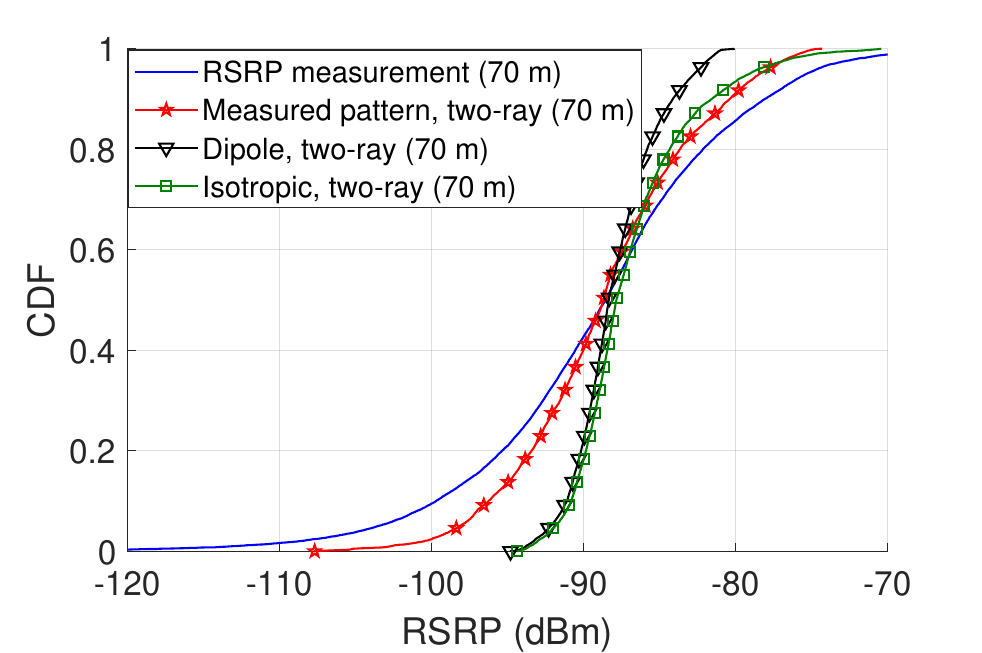}\label{fig:RSRP_cdf_50}}~
        \subfloat[CDF of RSRP, 110~m.]{\includegraphics[width=0.33\textwidth]{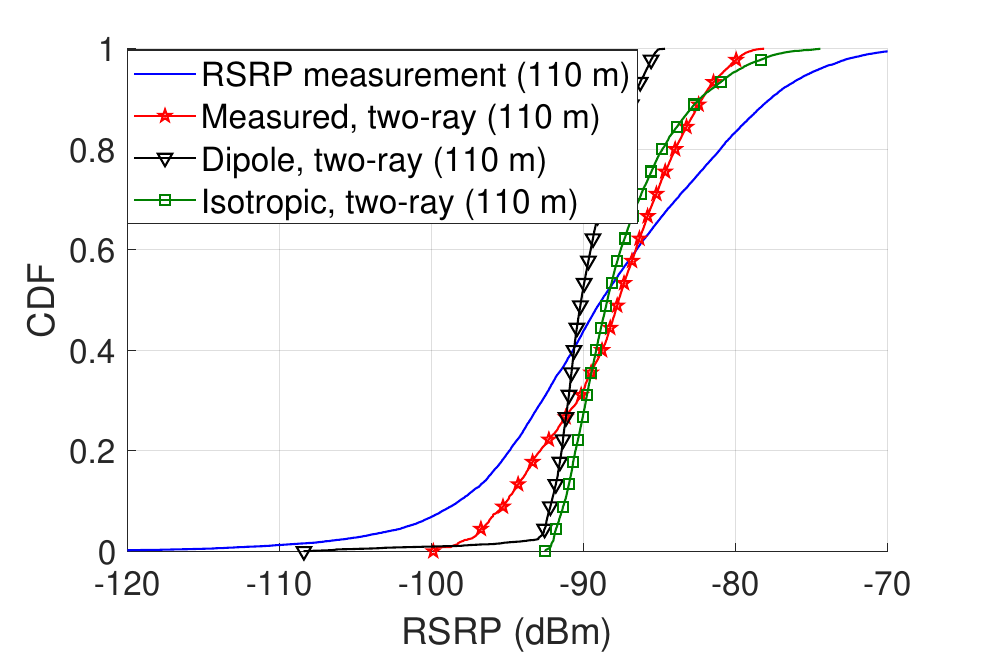}\label{fig:RSRP_cdf_110}}
        \vspace{-0.02in}
        \subfloat[CDF of fitting error between measured RSRP and path loss models, 30~m.]{\includegraphics[width=0.33\textwidth]{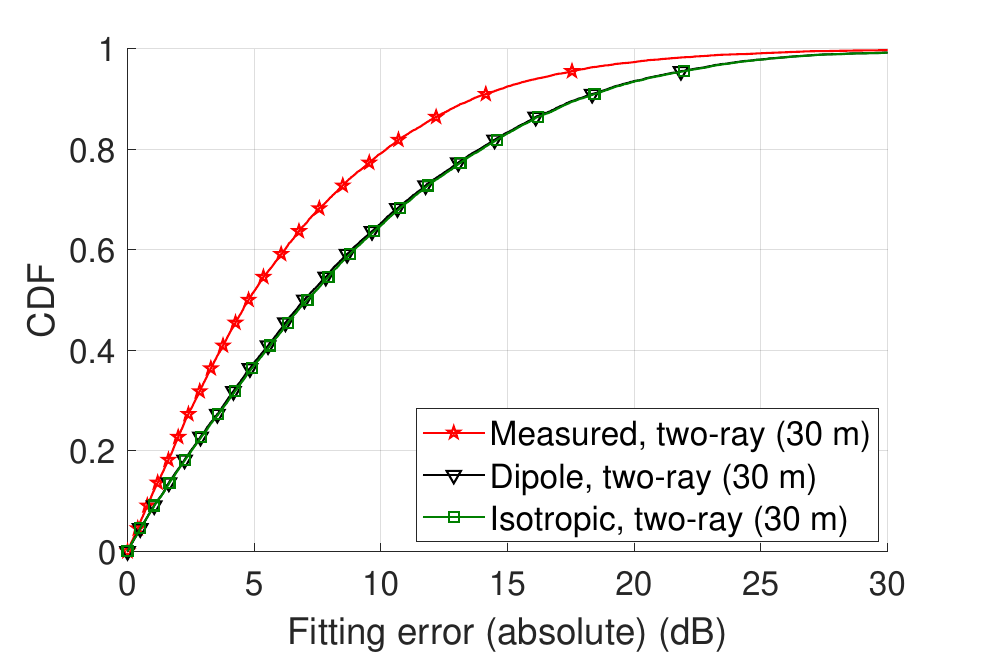}\label{fig:err_cdf_30}}~
	\subfloat[CDF of fitting error between measured RSRP and path loss models, 70~m.]{\includegraphics[width=0.33\textwidth]{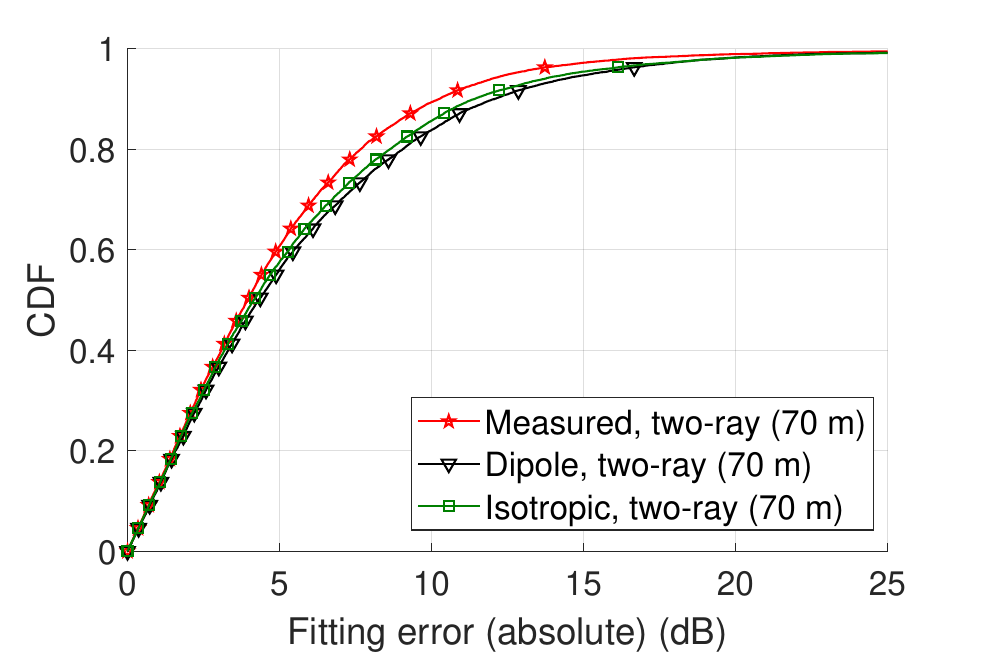}\label{fig:err_cdf_70}}~
        \subfloat[CDF of fitting error between measured RSRP and path loss models, 110~m.]{\includegraphics[width=0.33\textwidth]{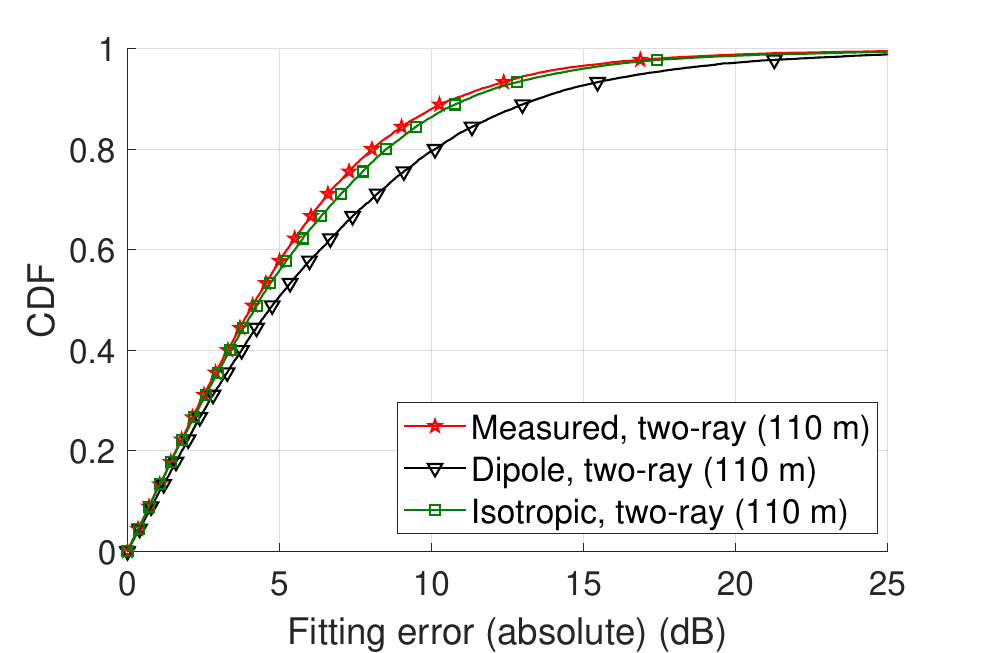}\label{fig:err_cdf_110}}
	\caption{CDF of RSRP and the fitting error to observe the similarity between measured RSRP and analytical RSRP. All three different antenna pattern options are considered.}\label{fig:RSRP_cdf}
\end{figure*}

\subsection{3D Antenna Radiation Pattern}\label{sec:3D_ant_rad}
In this paper, we investigate the importance of accurate 3D antenna radiation patterns in UAV air-to-ground propagation models. While terrestrial networks typically rely on isotropic antenna gain, which adequately captures the impact of omni-directional antenna patterns in the azimuth angle domain, this approach may overlook the elevation angle-dependent antenna gain in the 3D topology of air-to-ground networks. 

To observe the impact of different 3D antenna patterns, we consider three options:  1) a measured antenna pattern, 2) a dipole antenna pattern, and 3) a isotropic antenna pattern. In the first option of the measured antenna pattern, we utilize a measured 3D antenna pattern of Tx antenna (RM-WB1-DN) obtained from an anechoic chamber facility, available in IEEE DataPort~\cite{IEEEDataPort}, as well as an Rx antenna (SA-1400-5900) pattern from the specification sheet provided by the vendor~\cite{sa_1400}. Note that the datasheet of the Tx antenna pattern provided by the vendor~\cite{rm_wb1} includes data only for a limited number of frequencies, not including the 3.5~GHz spectrum used in this paper. Moreover, the datasheet considers only a single cross-section of the elevation angle, even though the antenna pattern is sensitive to the frequency and it may vary depending on the azimuth angle.

For the second option, we adopt the elevation-angle dependent donut shape dipole antenna pattern expression for both Tx and Rx antennas, given as~\cite{maeng2020interference,balanis2016antenna}
\begin{align}\label{eq:dipole_formula}
\mathsf{G}_{\rm Tx}(\theta)&=\mathsf{G}_{\rm Rx}(\theta)=\frac{\cos\left(\frac{\pi}{2}\cos\theta\right)}{\sin\theta}.
\end{align}
Fig.~\ref{fig:dipole_pat} shows the dipole antenna pattern obtained using the analytical dipole expression in \eqref{eq:dipole_formula}. The pattern exhibits the characteristic donut shape associated with a typical dipole antenna. In the isotropic antenna pattern setup, we utilize the isotropic antenna gain for both Tx and Rx antennas.

Fig.~\ref{fig:chamber} shows the setup for measuring the 3D antenna radiation pattern by using an anechoic chamber located at the Wireless Research Center (WRC), Wake Forest, NC. The Tx antenna is firmly installed at the right center of the chamber with styrofoam-type supports on the turntable, indicated by the intersection of laser lines. During the rotation of the turntable throughout a full circle, spherical measurement antennas capture the antenna gain across all elevation angles, enabling the generation of full 3D antenna patterns. The resulting pattern from the measurement is shown in Fig.~\ref{fig:Tx_ant_pat} by the Cartesian coordinates, Fig.~\ref{fig:Tx_ant_pat_2} by using the azimuth and elevation angles coordinates, and Fig.~\ref{fig:Tx_ant_pat_elev} by the elevation angle domain respectively. It is evident that the Tx antenna pattern is not the standard omni-directional dipole antenna pattern but rather a highly non-uniform directional antenna pattern in the 3D domain. It implies that neither the isotropic antenna gain nor the doughnut shape dipole antenna pattern will accurately capture the antenna effects in air-to-ground signal propagation. Moreover, a single cross-section of the elevation angle in \cite{rm_wb1} cannot fully capture the 3D antenna pattern. Fig.~\ref{fig:Rx_ant_pat} shows the antenna pattern in the elevation angle domain for the Rx antenna, as specified in the datasheet. It is observed that the Rx antenna pattern has the typical doughnut shape characteristic of a dipole pattern. Fig.~\ref{fig:Rx_ant_pat_mea} shows the Rx antenna pattern measured by the anechoic chamber in the elevation angle domain. We describe the mean, maximum, and minimum antenna gain of individual elevation angles with respect to the different azimuth angle cross-sections. It is observed that the overall antenna pattern fairly coincides with the antenna
pattern in the datasheet. Since this data for the Rx antenna was reasonably consistent with the measurements from our anechoic chamber and the antenna pattern is not very sensitive to the radio frequency from $1.4$ - $5.8$~GHz, we decided to use the antenna pattern of $2.4$~GHz frequency from the datasheet for the Rx antenna. Note that as the Rx antenna pattern is considered omni-directional in the azimuth angle domain, a single-cut elevation domain antenna pattern can cover the 3D directional pattern.

\section{Analysis of Impact of 3D Antenna Patterns on Air-to-ground Radio Propagation}\label{Sec:anal_ant_pat_a2g}

In this section, we assess the impact of the three different 3D antenna patterns mentioned in Section~\ref{sec:3D_ant_rad} on 3D path loss modeling. In particular, we compare the RSRP values obtained by collected datasets with those derived from the analytical path loss models combined with 3D antenna patterns. To examine the diverse effects of the antenna radiation pattern, we present a comparison of the measured and analytically derived RSRPs in a single figure across different domains, including time, distance, and elevation angle. Note that due to space limitation concerns, we present selected results from three different UAV altitudes (30~m, 70~m, and 110~m) even though we collected the datasets at five different UAV altitudes. 

\subsection{Time, Distance, and Elevation Angle Domains Analysis}\label{sec:t_d_e_domain}

Fig.~\ref{fig:RSRP_t} shows the RSRP change in the time domain. We overlap the RSRP calculated by the analytical model on top of the RSRP obtained from the measurement, which shows how much the analytical results follow the measurement RSRP. In the analytical models, we individually consider three different antenna patterns. From the first column's subfigures to the third column's subfigures (to the right), the height of the UAV increases from $30$~m to $110$~m, and from the first row's subfigures to the third row's subfigures (downward), the antenna pattern setups are changed. In particular, ``Measured antenna pattern'' refers to the antenna patterns obtained in the anechoic chamber measurements shown in Fig.~\ref{fig:ant_pat}, the dipole antenna pattern is the analytical model in~\eqref{eq:dipole_formula}, and isotropic antenna pattern considers a constant 3D pattern. The measured antenna pattern follows the fluctuation of the RSRP due to the zig-zag pattern of the trajectory for all three different heights in Fig.~\ref{fig:RSRP_t_30_1}, Fig.~\ref{fig:RSRP_t_70_1}, Fig.~\ref{fig:RSRP_t_110_1}, while the dipole and the isotropic antenna pattern do not fully capture the fluctuation of the RSRP, especially the deep fades and the peaks in Fig.~\ref{fig:RSRP_t_30_2}, Fig.~\ref{fig:RSRP_t_70_2}, Fig.~\ref{fig:RSRP_t_110_2}, Fig.~\ref{fig:RSRP_t_30_3}, Fig.~\ref{fig:RSRP_t_70_3}, Fig.~\ref{fig:RSRP_t_110_3}. This implies that the isotropic or simple dipole antenna pattern models are not sufficient to realistically characterize air-to-ground propagation for the considered scenario. Moreover, it is critical to utilize the accurate antenna pattern in modeling the path loss. We also observe that the two-ray path loss model provides a more accurate representation of the deep fading phenomenon in RSRP compared with the free space path loss model. In addition, the deep fades of the RSRP are not captured by the isotropic antenna pattern while both deep fading and the peak of RSRP are not captured by the dipole antenna pattern in Fig.~\ref{fig:RSRP_t_110_2}, Fig.~\ref{fig:RSRP_t_110_3}. This implies that the antenna gain of a low elevation angle (deep fading) is overestimated in both the isotropic and the dipole antenna patterns while the antenna gain of a high elevation angle (peak) is underestimated in the dipole antenna pattern.

Although time-domain RSRP figures compare the analytical RSRP with different antenna patterns with the measured RSRP, those results do not provide the full picture of the impact of the 3D antenna radiation patterns. For instance, distance domain RSRP figures can provide additional information on how the average signal power changes as a function of the distance. In particular, the variation of the RSRP with distance depends on the specific altitude and it is not monotonic -- RSRP gets maximized at certain distances for some altitudes. This is because the path loss and antenna patterns jointly affect the RSRP and the specific behavior depends on the 3D location of the UAV. Similarly, elevation angle domain RSRP figures can give insights into how the elevation angle affects the RSRP for different scenarios -- again, the specific behavior can be non-monotonic depending on the altitude of the UAV. Furthermore, the cumulative distribution functions (CDFs) can be used to analyze the distribution of RSRP and the model fitting error for different altitudes, which can provide better numerical comparisons between measurement and analysis when compared with what can be extracted from the previous figures.

Motivated by these, we provide several additional results. Fig.~\ref{fig:RSRP_d} shows the RSRP versus distance for three different antenna models and three different altitudes. As was also observed in the results in Fig.~\ref{fig:RSRP_t}, we observe that the path loss model that uses the measured antenna pattern results in the closest match with the measured RSRP for all heights. In the dipole antenna pattern with 70~m UAV height, the RSRP is underestimated/overestimated in the short/long distance. In the isotropic antenna pattern, the slope of the RSRP with the free space path loss model is isotropic since the path loss becomes only a function of distance in \eqref{eq:PL_fs}. However, the slope does not match with the measured RSRP as the distance increases.

Finally, Fig.~\ref{fig:RSRP_e} shows the measurement and analytical RSRP depending on the elevation angle. In these results, we can clearly observe that the path loss model that uses the measured antenna pattern matches best with the measured RSRP in the elevation domain when compared to the other two antenna pattern models. In addition, the slope of the analytical RSRP in the dipole antenna pattern exhibits the poorest match with the corresponding measurement.

\subsection{CDF Analysis of RSRP and Modeling Error}
In Fig.~\ref{fig:RSRP_cdf}a-Fig.~\ref{fig:RSRP_cdf}c, we first show the CDF of analytically derived RSRP for different antenna models at different altitudes and compare them with the measured RSRP. Subsequently, in Fig.~\ref{fig:RSRP_cdf}d-Fig.~\ref{fig:RSRP_cdf}e we analyze the CDFs of the modeling error for the three different antenna models. We observe that the CDF of RSRP obtained from the measured antenna pattern is the closest to the CDF of the measured RSRPs for all heights. In addition, the two-ray path loss model results in a CDF that is similar to the CDF of the free-space model -- this is consistent with the RSRP results for the two models in Fig.~\ref{fig:RSRP_t}, Fig.~\ref{fig:RSRP_d}, Fig.~\ref{fig:RSRP_e}. 

The fitting errors in Fig.~\ref{fig:err_cdf_30}, Fig.~\ref{fig:err_cdf_70}, Fig.~\ref{fig:err_cdf_110} are calculated by the absolute difference between the measurement RSRP and the analytical RSRP. We observe that the fitting error is the smallest for the measured antenna pattern for all heights, and the gap between the measured antenna pattern and other antenna patterns is the largest at 30~m height.

\begin{figure}[t!]
	\centering
	\subfloat[Measured total antenna pattern (Tx+Rx) from the antenna patterns in Fig.~\ref{fig:ant_pat}.] {\includegraphics[width=0.35\textwidth]{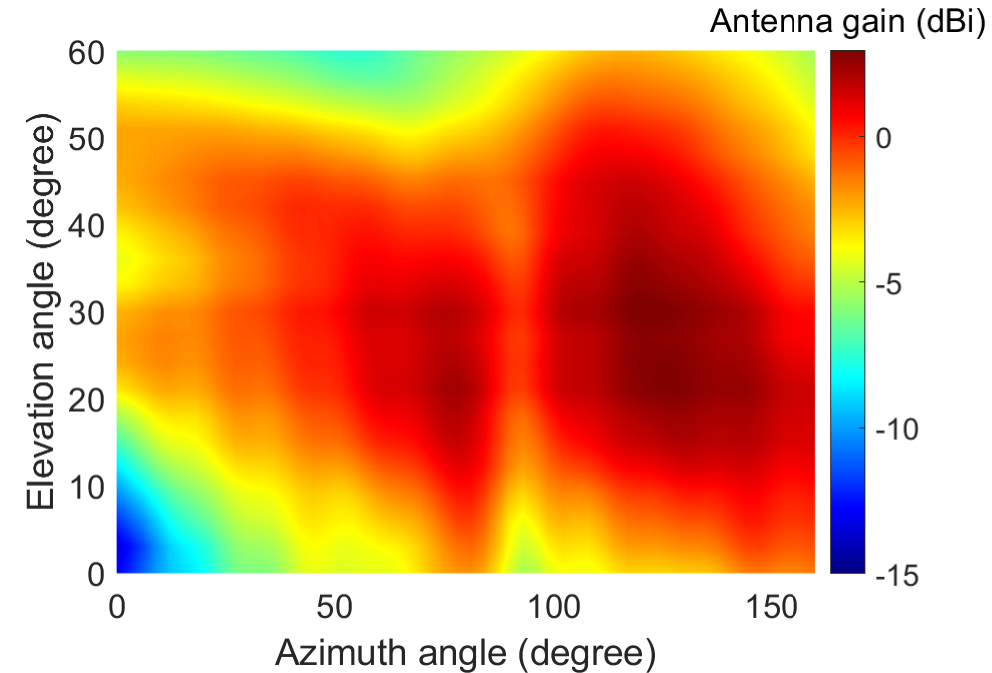}\label{fig:ant_tot_mea}}
        \vspace{-0.02in}
        \subfloat[Estimated total antenna pattern (Tx+Rx) from the RSRP and GPS log datasets of all UAV heights.]{\includegraphics[width=0.35\textwidth]{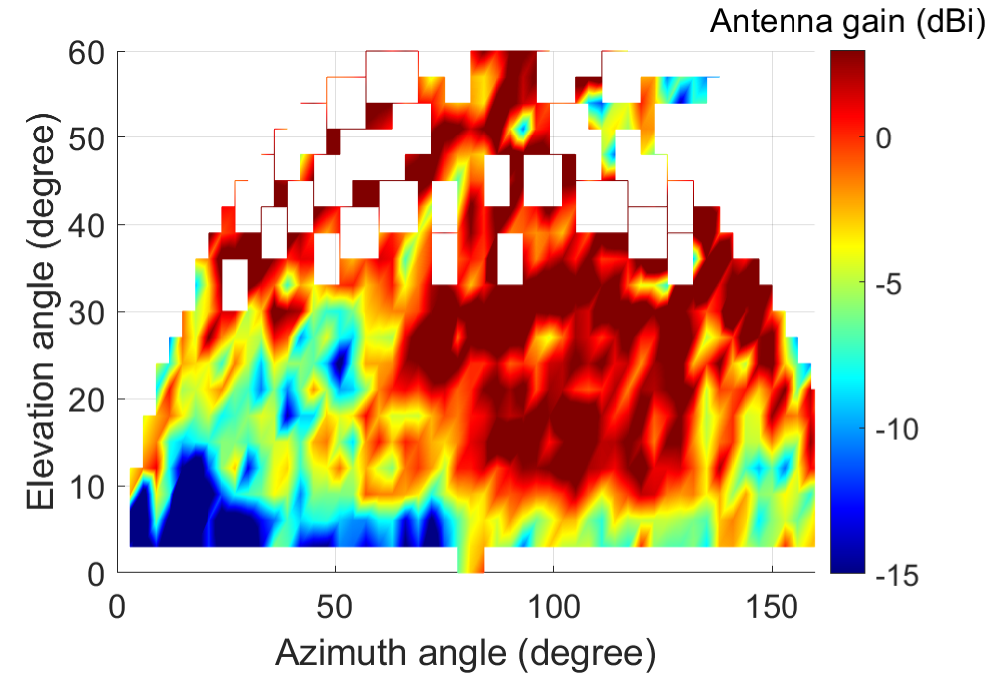}\label{fig:ant_tot_est}}
        \vspace{-0.02in}
        \subfloat[Error between measured and estimated total antenna pattern.]{\includegraphics[width=0.35\textwidth]{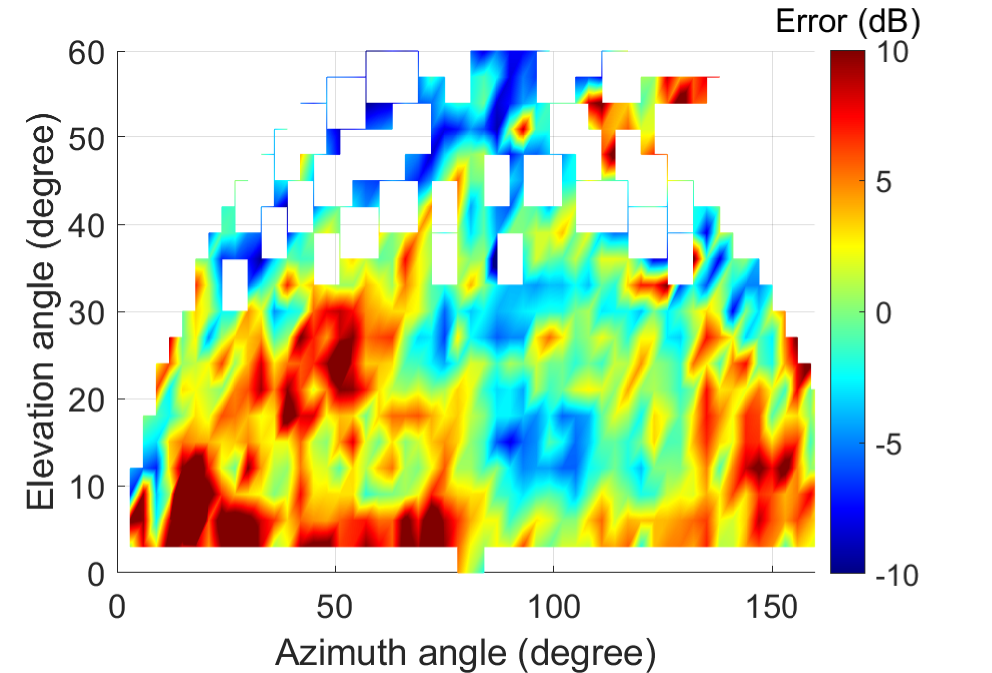}\label{fig:ant_tot_err}}
        \vspace{-0.02in}
        \subfloat[Number of RSRP samples ($N_{\rm xy}$).]{\includegraphics[width=0.35\textwidth]{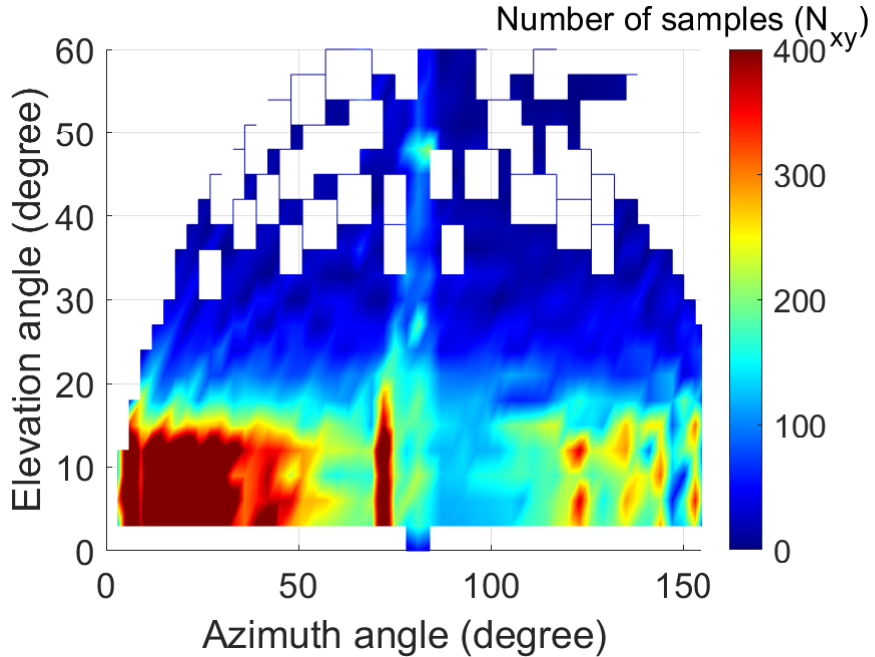}\label{fig:ant_tot_Nxy}}
	\caption{Comparison of (a) measured antenna pattern with (b) the estimated antenna pattern. (c) The error of the antenna pattern is generated by subtracting (b) the estimated antenna pattern from (a) the measured antenna pattern. (d) the number of RSRP samples ($N_{\rm xy}$) located at the corresponding azimuth and elevation angle is used in generating the estimated antenna pattern in (b).}\label{fig:ant_tot}
\end{figure}

\subsection{Antenna Pattern Estimation}
For the antenna pattern analysis in the earlier sections, we use 3D antenna models together with corresponding path loss models, derive an analytical RSRP for the corresponding joint model, and finally, compare the measured RSRP with the analytical RSRP. In this subsection, we take a reverse approach:  we estimate the 3D antenna pattern from the RSRP measurements and compare the estimated antenna pattern with the measured antenna pattern from the anechoic chamber (see Fig.~\ref{fig:ant_pat}). Fig~\ref{fig:ant_tot_mea} shows the combined total antenna pattern of the Tx and Rx with respect to the elevation and azimuth angles. To obtain the combined antenna pattern, we utilize the Tx antenna pattern shown in Fig.~\ref{fig:Tx_ant_pat_2} and the Rx antenna pattern shown in Fig.~\ref{fig:Rx_ant_pat}. 

In Fig.~\ref{fig:ant_tot_est}, we present the estimated 3D antenna pattern in the azimuth and the elevation angle domains. To obtain this estimated antenna pattern, we utilize all available UAV height datasets, including 30 m, 50 m, 70 m, 90 m, and 110 m. Since we have access to the 3D location of the UAV for each RSRP measurement through GPS logs, we can determine the azimuth and elevation angles of the LoS between the BS tower and the UAV. By subtracting transmit power and the free space path loss component without antenna patterns from the RSRP value, we can obtain the estimated 3D antenna pattern, which is given by
\begin{align}\label{eq:ant_pat_est}
    &\hat{\mathsf{G}}_{\rm tot}(\phi_{\rm x},\theta_{\rm y})\nonumber\\
    &=\frac{1}{N_{\rm x\rm y}}\sum_{i=1}^{N_{\rm x\rm y}}\left[\mathsf{RSRP}^i(\phi_{\rm x},\theta_{\rm y})-\mathsf{P}_{\rm Tx}+\mathsf{PL}_{\rm FS}^i(\phi_{\rm x},\theta_{\rm y})\right],
\end{align}
where $N_{\rm x\rm y}$ denotes the number of RSRP samples where the UAV is located at a LoS azimuth angle of $\phi_{\rm x}$ and a LoS elevation angle of $\theta_{\rm y}$, and $\mathsf{PL}_{\rm FS}^i$ indicates free space path loss in \eqref{eq:PL_fs} with $\mathsf{G}_{\rm Tx}(\phi_{\rm x},\theta_{\rm y})=\mathsf{G}_{\rm Rx}(\phi_{\rm x},\theta_{\rm y})=1$. 

Note that the estimated antenna pattern may have limitations in terms of accuracy, particularly when it comes to removing the effects of shadowing and ground-reflected signals. These limitations become more apparent when the number of samples from the measurements is insufficient. From Fig.~\ref{fig:ant_tot_est}, it is observed that the estimated antenna pattern occupies a restricted angle space that aligns with the trajectory of the flight. It spans approximately from $5$ to $160$ degrees in the azimuth angle range and from $5$ to $60$ degrees in the elevation angle range. The white spaces in the figure occur because they correspond to angles that fall outside the range of the UAV's flying area. In such cases, we can easily interpolate the unknown antenna patterns to estimate their characteristics, which will be studied in future work. Comparing with the measured antenna pattern in Fig.~\ref{fig:ant_tot_mea}, we observe an overall similarity in terms of directivity. High antenna gain is observed in the azimuth angle range of $60$ to $140$ degrees and the elevation angle range of $10$ to $40$ degrees. However, the accuracy of the estimated antenna pattern degrades in the low elevation angle range ($<10$) since the impact of scattering is relatively high and RSRP-based estimation achieves higher error. In addition, the accuracy of the estimated antenna pattern is low in certain higher elevation angle ranges ($>20$) due to the limited number of samples available in the measurement datasets.

Fig.~\ref{fig:ant_tot_err} shows the relative error between the measured antenna pattern in Fig.~\ref{fig:ant_tot_mea} and the estimated antenna pattern in Fig.~\ref{fig:ant_tot_est}. We observe that the mismatch between the two antenna patterns is relatively low within certain angle ranges. Specifically, the azimuth angle range of $60$ to $140$ degrees and the elevation angle range of $10$ to $40$ degrees exhibit a low level of mismatch. On the other hand, the mismatch between the antenna patterns is high within some other angle ranges. In particular, the azimuth angle range of $10$ to $80$ degrees and the elevation angle range of $5$ to $10$ degrees exhibit a higher level of mismatch.

Fig.~\ref{fig:ant_tot_Nxy} shows the number of RSRP samples at the corresponding azimuth and elevation angles ($N_{\rm xy}$) utilized in estimating the 3D antenna pattern in \eqref{eq:ant_pat_est}. It is observed that the number of samples decreases as the elevation angle increases. In particular, the number of samples dramatically decreases from $15$ degrees, and after $20$ degrees, the number of samples mostly becomes less than $100$.

\section{Localization of BSs by Using UAVs}\label{sec:localization}
In this section, we focus on the problem of localizing the source BS by using the collected RSRP data over the trajectory of the UAV. In particular, we propose to use our air-to-ground propagation models and 3D antenna patterns to improve localization accuracy. Note that while the localization of a ground BS by a UAV may have its own merits in a cellular-connected UAV network (e.g. for a UAV to stay within the coverage of its serving BS during a flight~\cite{bulut2018trajectory}), the proposed solution is generic and can also be applied in other scenarios where a UAV aims to localize a different ground signal source, e.g. for search and rescue operations.

\subsection{RSRP-based Least Square Estimation}
The 3D location of a BS and a UAV can be represented by
\begin{align}
    \mathbf{l}^{\rm bs}&=(\omega^{\rm bs},\psi^{\rm bs},h^{\rm bs}),\;\mathbf{l}^{\rm uav}_i=(\omega^{\rm uav}_i,\psi^{\rm uav}_i,h^{\rm uav}_i),
\end{align}
where $\omega$, $\psi$, and $h$ denote the longitude, latitude, and altitude of the BS location, which can be obtained by the GPS sensors of the UAV. The discrete-time location of the UAV during the flight is indicated by $\mathbf{l}^{\rm uav}_i$. The horizontal distance between a BS and a UAV can be expressed as
\begin{align}\label{eq:dh}
    &d_{\rm h,i}=\arccos\left(\sin\psi^{\rm uav}_i\sin\psi^{\rm bs}+\cos\psi^{\rm uav}_i\right.\nonumber\\
    &\left.\cos\psi^{\rm bs}\cos(\omega^{\rm bs}-\omega^{\rm uav}_i)\right){\rm R},\nonumber\\
    &\approx\sqrt{(\omega^{\rm bs}-\omega^{\rm uav}_i)^2\cos(\psi_0)^2+(\psi^{\rm bs}-\psi^{\rm uav}_i)^2}\frac{{\rm R}\pi}{180},
\end{align}
where ${\rm R}\approx6378137$~m denotes the radius of earth. The approximation in \eqref{eq:dh} comes from the equirectangular projection, transforming spherical coordinates into planar coordinates~\cite{snyder1997flattening}, and $\psi_0$ denotes the standard parallels decided by the scale of the projection. Note that since the size of the area of the experiment site is small, the approximation is sufficiently accurate. The vertical distance between a BS and a UAV is given by $d_{\rm v}=|h^{\rm bs}-h^{\rm uav}_i|$. Then, the 3D distance ($d_3$) between a BS and a UAV can be calculated by
\begin{align}\label{eq:d3_1}
    d_{3,i}^2&=d_{\rm h,i}^2+d_{\rm v}^2.
\end{align}

From \eqref{eq:received} and \eqref{eq:PL_fs}, the received signal strength (RSRP) can be reformulated as
\begin{align}
    r_i&=\frac{\mathsf{P}_{\rm Tx}\lambda^2}{(4\pi)^2}\frac{\mathsf{G}_{\rm Tx,i}\mathsf{G}_{\rm Rx,i}}{d_{{\rm 3},i}^2}+\Tilde{s},
\end{align}
where $\Tilde{s}$ indicates additive components including shadowing and ground reflected signal. Then, by neglecting the additive components $\Tilde{s}$, we can represent the distance as follow:
\begin{align}\label{eq:d_by_FS}
    d_{\rm 3,i}^2&=\frac{\mathsf{P}_{\rm Tx}\lambda^2}{(4\pi)^2}\frac{\mathsf{G}_{\rm Tx,i}\mathsf{G}_{\rm Rx,i}}{r_i}.
\end{align}

To formulate the least square (LS) estimator using multiple RSRP measurements over the trajectory of the UAV, we first subtract the square of the 3D distance at a reference UAV location ($r$) from all other UAV locations ($i$)~\cite{guvenc2012fundamental,kwon2023rf}, which can be written as follows: 
\begin{align}\label{eq:d_sub_d}
    &d_{{\rm 3},r}^2-d_{{\rm 3},i}^2=\left(\frac{{\rm R}\pi}{180}\right)^2\left\{(\omega^{\rm bs}-\omega^{\rm uav}_{r})^2\cos(\psi_0)^2+\right.\nonumber\\
    &\left.(\psi^{\rm bs}-\psi^{\rm uav}_{r})^2-(\omega^{\rm bs}-\omega^{\rm uav}_{i})^2\cos(\psi_0)^2-(\psi^{\rm bs}-\psi^{\rm uav}_{i})^2\right\},
\end{align}
where $d_{{\rm 3},r}$, $\psi^{\rm uav}_{r}$, $\omega^{\rm uav}_{r}$ denote the 3D distance and the coordinates of the reference point, respectively, and $i=1,\cdots,N$ where $N$ is the number of UAV locations where the RSRPs will be used for localization, excluding the reference location $r$. By reorganizing  \eqref{eq:d_sub_d} in matrix form for $i=1,\cdots,N$,  we can obtain the following linear expression:
\begin{align}\label{eq:LS_sol}
    \mathbf{A}\hat{\mathbf{l}}=\mathbf{B},
\end{align}
where $\mathbf{A}$, $\mathbf{B}$, $\hat{\mathbf{l}}$ are given at the top of next page in \eqref{eq:A_B_l}. After simplifying \eqref{eq:d_sub_d} and substituting $d_{{\rm 3},r}^2$ and $d_{{\rm 3},i}^2$ into \eqref{eq:d_by_FS}, we can obtain $\mathbf{A}$, $\mathbf{B}$. Finally, we can obtain the estimated location of the BS (longitude and latitude) $\hat{\mathbf{l}}$ by multiplying both sides of \eqref{eq:LS_sol} with the Moore-Penrose inverse of $\mathbf{A}~(\mathbf{A}^{\dagger})$, which is given by
\begin{align}\label{eq:LS_sol_final}
    \hat{\mathbf{l}}=\mathbf{A}^{\dagger}\mathbf{B}.
\end{align}

Note that the localization accuracy depends critically on the RSRPs used at specific locations over the trajectory of the UAV, i.e., the set of $\mathbf{l}^{\rm uav}_i$ for $i=1,\cdots,N$. In this work, we consider both random and more structured approaches of selecting $\mathbf{l}^{\rm uav}_i$ over the UAV's trajectory, and compare their localization performance. 

\begin{figure*}[t]
\begin{align}\label{eq:A_B_l}
 \mathbf{A}&=\left[ \begin{array}{ccc}
    (-2\omega^{\rm uav}_{r}+2\omega^{\rm uav}_{1})\cos(\psi_0)^2& \cdots & (-2\omega^{\rm uav}_{r}+2\omega^{\rm uav}_{N})\cos(\psi_0)^2\\
    -2\psi^{\rm uav}_{r}+2\psi^{\rm uav}_{1}& \cdots &-2\psi^{\rm uav}_{r}+2\psi^{\rm uav}_{N}
    \end{array}\right]^{\rm T}\nonumber\\
\mathbf{B}&=\left[ \begin{array}{c}
    \left(\frac{180}{{\rm R}\pi}\right)^2\frac{\mathsf{P}_{\rm Tx}\lambda^2}{(4\pi)^2}(\frac{\mathsf{G}_{\rm bs,r}\mathsf{G}_{\rm uav,r}}{r_r}-\frac{\mathsf{G}_{\rm bs,1}\mathsf{G}_{\rm uav,1}}{r_1})-(\omega^{\rm uav}_{r})^2+(\omega^{\rm uav}_{1})^2+(-(\omega^{\rm uav}_{r})^2+(\omega^{\rm uav}_{1})^2)\cos(\psi_0)^2\\
    \vdots\\
    \left(\frac{180}{{\rm R}\pi}\right)^2\frac{\mathsf{P}_{\rm Tx}\lambda^2}{(4\pi)^2}(\frac{\mathsf{G}_{\rm Tx,r}\mathsf{G}_{\rm Rx,r}}{r_r}-\frac{\mathsf{G}_{\rm Tx,N}\mathsf{G}_{\rm Rx,N}}{r_N})-(\omega^{\rm uav}_{r})^2+(\omega^{\rm uav}_{N})^2+(-(\omega^{\rm uav}_{r})^2+(\omega^{\rm uav}_{N})^2)\cos(\psi_0)^2
    \end{array}\right]\nonumber\\
\hat{\mathbf{l}}&=\left[ \begin{array}{cc}
    \hat{\omega}^{\rm bs}&\hat{\psi}^{\rm bs}
    \end{array}\right]^{\rm T}.
\end{align}
\hrulefill
\end{figure*}

\subsection{Fixed-point Iterative Algorithm}
The LS estimator solution for the localization problem in \eqref{eq:LS_sol_final} requires the antenna gains $\mathsf{G}_{\rm Tx}$, $\mathsf{G}_{\rm Rx}$ for constructing $\mathbf{B}$. However, the antenna gains are also the function of the location of the BS $\omega^{\rm bs}$, ${\psi}^{\rm bs}$, which is what we are aiming to estimate. Therefore, we propose a fixed-point iterative algorithm to solve for the BS's location, where: 1) the LS estimator is solved by given antenna gains; 2) the antenna gain is updated by the estimated BS location; and finally, 3) the estimated BS location is re-updated by the LS estimator with the updated antenna gain. We initialize the antenna gain as $1$ for all UAV locations (isotropic antenna pattern) and we utilize the given antenna patterns (measured, dipole, isotropic) from the second iteration. In this way, we take into account the impact of the antenna pattern to improve the localization performance. The detailed steps are described in Algorithm~\ref{algorithm:localization}. We denote antenna patterns as $\mathsf{G}_{\rm Tx}^{\rm pat}(\phi,\theta)$, $\mathsf{G}_{\rm Rx}^{\rm pat}(\phi,\theta)$.

\begin{algorithm}[t!]
  \caption{Fixed-point Iterative Algorithm for Localization}
  \label{algorithm:localization}
    \begin{algorithmic}[1]
        \State \textbf{Input:} $r_i$, $\omega^{\rm uav}_{i}$, $\psi^{\rm uav}_{i}~\forall i$, $\mathsf{P}_{\rm Tx}$, $\mathsf{G}_{\rm Tx}^{\rm pat}(\phi,\theta)$, $\mathsf{G}_{\rm Rx}^{\rm pat}(\phi,\theta)$
        \State \textbf{Initialize:} $\mathsf{G}_{\rm Tx,i}^{(1)}=1$, $\mathsf{G}_{\rm Rx,i}^{(1)}=1~\forall i$
        \State \textbf{Iterations}:
        \For{$k = 1, \dots , K$}
        \State Calculate $\mathbf{A}^{(k)}$, $\mathbf{B}^{(k)}$ by \eqref{eq:A_B_l}
        \State Obtain $\hat{\mathbf{l}}^{(k)}$ by solving LS estimator in \eqref{eq:LS_sol}
        \State Update $\mathsf{G}_{\rm Tx,i}^{(k+1)}$, $\mathsf{G}_{\rm Rx,i}^{(k+1)}~\forall i$ from the estimated location of the BS $\hat{\mathbf{l}}^{(k)}$, and antenna patterns $\mathsf{G}_{\rm Tx}^{\rm pat}(\phi,\theta)$, $\mathsf{G}_{\rm Rx}^{\rm pat}(\phi,\theta)$ 
        \EndFor
    \end{algorithmic}
\end{algorithm}

\subsection{Offline Localization}\label{sec:offline_local}
In this section, we describe the offline localization method using the Algorithm~\ref{algorithm:localization}. In the offline mode, we utilize the entire dataset after finishing the UAV flight to localize the BS tower location. For instance, the input dataset of Algorithm~\ref{algorithm:localization} $r_i$, $\omega^{\rm uav}_{i}$, $\psi^{\rm uav}_{i}$ can be randomly chosen $N$ samples out of the whole dataset where $i=1,~\cdots,~N$ to estimate the location of the BS tower $\hat{\mathbf{l}}^{\rm bs}$. The antenna patterns $\mathsf{G}_{\rm Tx}^{\rm pat}(\phi,\theta)$, $\mathsf{G}_{\rm Rx}^{\rm pat}(\phi,\theta)$ can be considered by the measured, dipole, and isotropic antenna patterns.

\subsection{Online Localization}\label{sec:online_local}
In this section, we propose a real-time algorithm for online localization by adopting the RSRP-based fixed-point iterative algorithm in Algorithm~1. As a UAV collects the received signal and obtains RSRP from the new measurement, it continuously updates the estimated location of the signal source in real time. This updating process leverages the estimated locations obtained from previous measurements, along with the new measurement, to refine the localization estimate. In the proposed online localization algorithm, we randomly pick $M$ RSRP samples out of the $N_{\rm buf}$ samples in the buffer of collected RSRP, which can be given by $\mathbf{r}_{\rm buf}=\{r_1,\cdots,r_{N_{\rm buf}}\}$ where the maximum size of buffer is $N_{\rm max}$. Besides, the coordinates of the UAV with the corresponding RSRPs are buffered in $\boldsymbol{\omega}_{\rm buf}=\{\omega_1,\cdots,\omega_{N_{\rm buf}}\}$, $\boldsymbol{\psi}_{\rm buf}=\{\psi_1,\cdots,\psi_{N_{\rm buf}}\}$. If the size of the buffer becomes full, the new RSRP replaces the most outdated RSRP. After that, we apply Algorithm~1 and obtain the estimated location $\hat{\mathbf{l}}$. Note that in the offline localization, $N$ samples are chosen from the entire dataset once, while in the online localization, $M$ samples are selected from the buffer for every data collection, and data collections are executed $J$ times.

We further consider a weighted localization algorithm where multiple location estimates are combined after being scaled with their normalized confidence weights. The motivation is that due to the non-uniform antenna patterns in 3D, the use of certain combinations of UAV locations for estimating the source location will be more favorable than other locations. Such weighted localization approaches have been used extensively in the literature for non-line-of-sight mitigation, see e.g.~\cite{guvenc2009survey} and the references therein. The confidence of each estimated location can be characterized based on the residual for that location estimate~\cite{chen1999non}, which can be expressed as
\begin{align}\label{eq:resi}
    e_{\rm res,j}&=\frac{1}{M}\sum_{i=1}^{M}|d_{3,i}-\hat{d}_{3,i}(\hat{\omega}^{\rm bs},\hat{\psi}^{\rm bs})|^2,
\end{align}
where $d_{3,i}$ comes from \eqref{eq:d_by_FS} and $\hat{d}_{3,i}(\hat{\omega}^{\rm bs},\hat{\psi}^{\rm bs})$ is calculated by \eqref{eq:dh}, \eqref{eq:d3_1}. Then, the weight of the estimated location can be written as follows: 
\begin{align}\label{eq:weight}
    \mu_j = \frac{N_{\rm buf}}{\log_{10}(e_{\rm res,j})}.
\end{align}
As the residual is smaller and the size of the buffer is larger, the weight of the estimated location is higher. The real-time estimated location of the signal source can be updated by the weighted linear combination, which is given by
\begin{align}
    \hat{\mathbf{l}}^{\star} = \frac{\sum_{k=1}^{j}\hat{\mathbf{l}}_k\mu_k}{\sum_{k=1}^{j}\mu_k}.
\end{align}
The detailed algorithm of the online localization is described in Algorithm~2. In the algorithm, $\mathbf{x}_{\setminus x_1}$ represents the operation that excludes $x_1$ entry from $\mathbf{x}$.

Instead of choosing $M$ samples randomly from the buffers in line 13 of Algorithm~\ref{algorithm:online}, we can also explore alternative strategies for sample selection. The first option can be the equal interval selection, which chooses samples with equal intervals from the buffer. Since measurements are periodically conducted with uniform time intervals, the selected samples have equal time intervals. Therefore, the index of $M$ samples from the equal interval selection can be $1, \lfloor\frac{N_{\rm buf}}{M}\rceil, \lfloor\frac{2N_{\rm buf}}{M}\rceil, \cdots, N_{\rm buf}$ where $\lfloor x \rceil$ denotes the round operation. 

The second option for sample selection is the nearby waypoints selection strategy. This approach involves utilizing samples that are in close proximity to the waypoints along the UAV's route. The waypoints serve as reference points that outline the trajectory of the UAV, allowing us to choose a set of samples that are spatially well separated. If the number of waypoints contained in the buffer is smaller than $M$,  we choose multiple samples that are near the same waypoint to fulfill the $M$ sample requirement. For example, consider that the buffer contains samples around waypoints 1, 2, 3, and 4, and we need $M=10$ samples. In this case, we pick 3 samples around each of the waypoints 1 and 2, and 2 samples around each of the waypoints 3 and 4.

\begin{algorithm}
  \caption{Online Localization}
  \label{algorithm:online}
    \begin{algorithmic}[1]
        \For{$j = 1, \dots , J$}
        \State \textbf{New dataset}: 
        \State Collect $r^{(j)}$, $\omega^{\rm uav(j)}$, $\psi^{\rm uav(j)}$  from the new measurement
        \State \textbf{Buffers update}:
        \If{$N_{\rm buf}< N_{\rm max}$} 
            \State $\mathbf{r}_{\rm buf}\gets[\mathbf{r}_{\rm buf},~r^{(j)}]$, $\boldsymbol{\omega}^{\rm uav}_{\rm buf}\gets[\boldsymbol{\omega}^{\rm uav}_{\rm buf},~\omega^{\rm uav(j)}]$,\\ $\boldsymbol{\psi}^{\rm uav}_{\rm buf}\gets[\boldsymbol{\psi}^{\rm uav}_{\rm buf},~\psi^{\rm uav(j)}]$
        \ElsIf{$N_{\rm buf}\geq N_{\rm max}$} 
            \State $\mathbf{r}_{\rm buf}\gets[\mathbf{r}_{\rm buf\setminus r_1},~r^{(j)}]$,\\ $\boldsymbol{\omega}^{\rm uav}_{\rm buf}\gets[\boldsymbol{\omega}^{\rm uav}_{\rm buf\setminus \omega_1},~\omega^{\rm uav(j)}]$, $\boldsymbol{\psi}^{\rm uav}_{\rm buf}\gets[\boldsymbol{\psi}^{\rm uav}_{\rm buf\setminus \psi_1},~\psi^{\rm uav(j)}]$
        \EndIf
        \State \textbf{Localization}:
        \State Randomly choose $M$ samples from $\mathbf{r}_{\rm buf}, \boldsymbol{\omega}^{\rm uav}_{\rm buf}, \boldsymbol{\psi}^{\rm uav}_{\rm buf}$ 
        \State And, obtain a estimated location $\hat{\mathbf{l}}^{(j)}$ from $M$ samples by applying Algorithm~1
        \State \textbf{Weighted linear combination}:
        \State Calculate the weight $\mu^{(j)}$ from \eqref{eq:resi}, \eqref{eq:weight}
        \State Update a real-time estimated location by $\hat{\mathbf{l}}^{\star(j)}=\frac{\sum_{k=1}^{j}\hat{\mathbf{l}}^{(k)}\mu^{(k)}}{\sum_{k=1}^{j}\mu^{(k)}}$
        \EndFor
    \end{algorithmic}
\end{algorithm}

\section{Numerical Results}\label{Sec:num_results_localization}

In this section, we evaluate the performance of the localization algorithms with different antenna patterns discussed in Section~\ref{sec:localization}, considering both the offline and the online approaches. In the offline mode, the BS tower is localized using the entire dataset, while real-time localization is examined in the online mode.

\subsection{Performance of Offline RSRP-based Localization}
\begin{figure}[t!]
	\centering
	\includegraphics[width=0.48\textwidth]{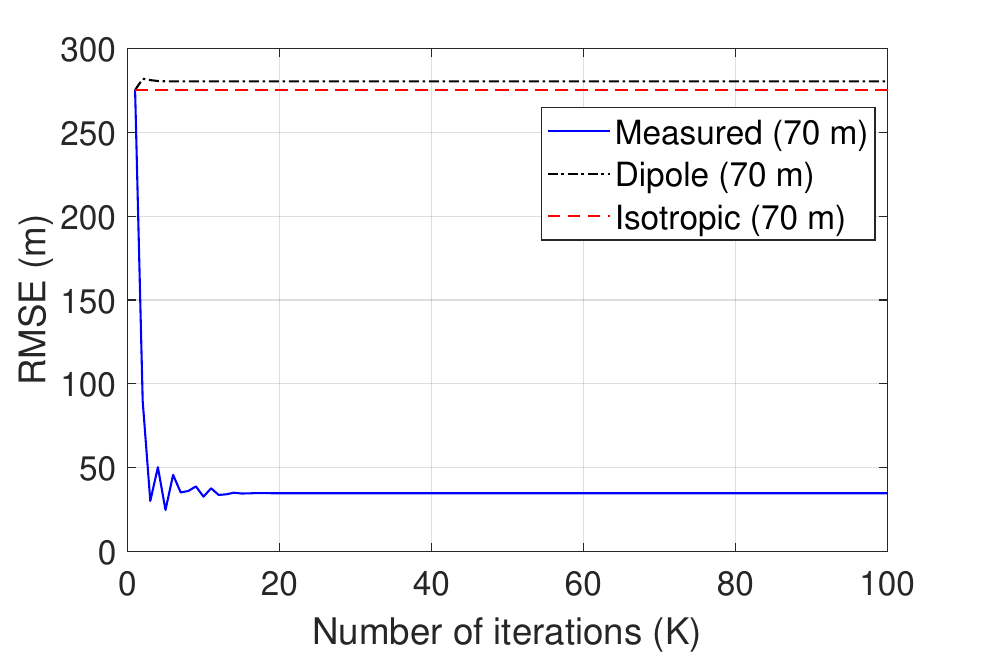}
	\caption{The convergence of Algorithm 1 with different antenna patterns where the number of samples $N=300$.}\label{fig:convergence}
\end{figure}
\begin{figure}[t!]
	\centering
	\subfloat[Convergence of RMSE at 70~m and 110~m UAV heights.]{\includegraphics[width=0.48\textwidth]{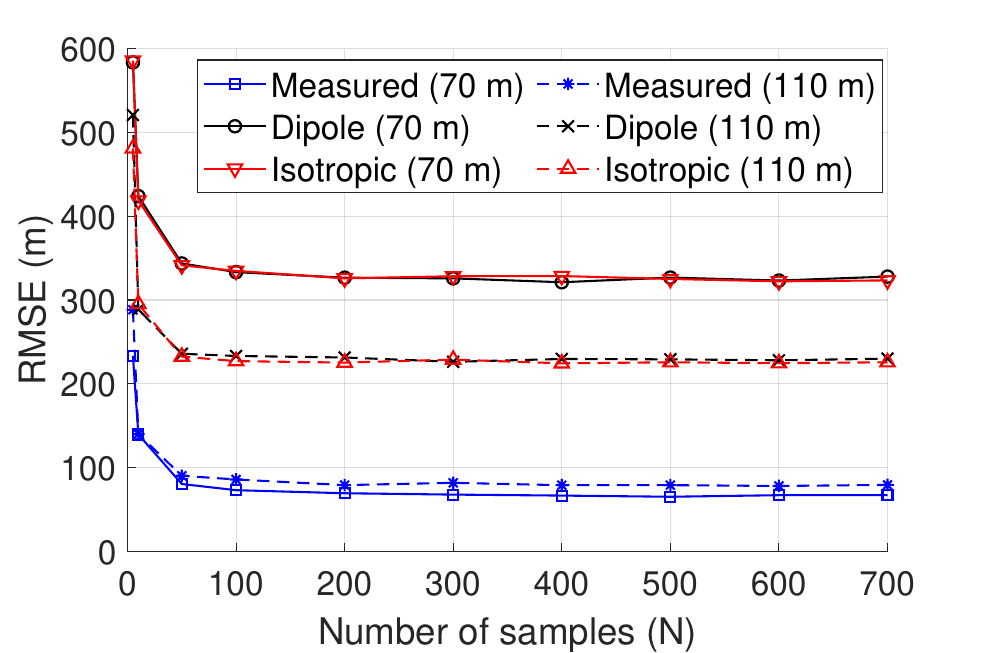}\label{fig:dist_err_70}}

        \subfloat[Converged RMSE vs. UAV height.]{\includegraphics[width=0.48\textwidth]{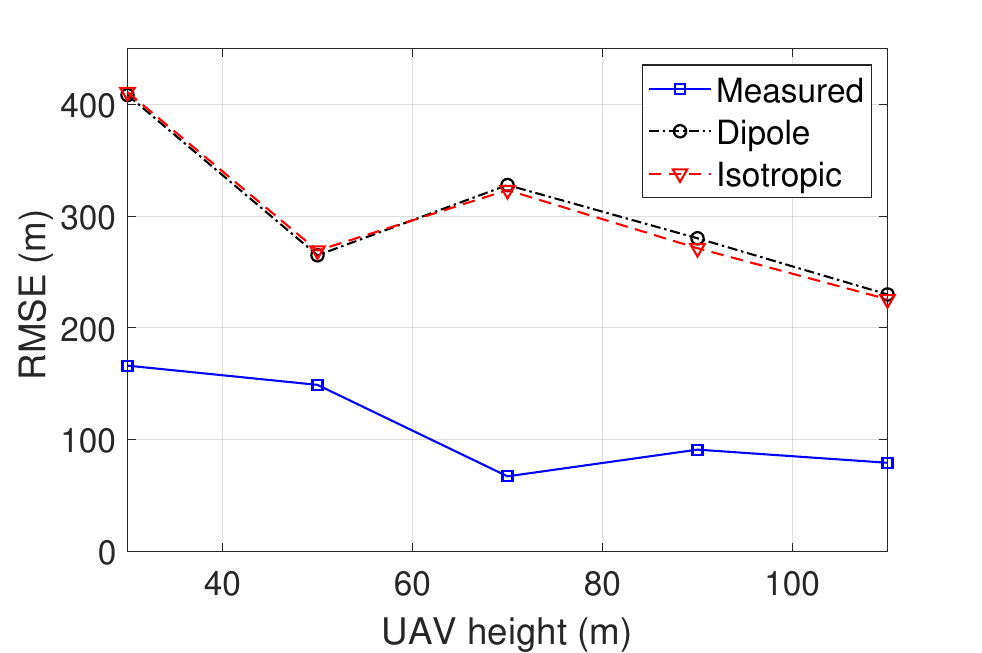}\label{fig:dist_err_all}}~
	\caption{(a) The RMSE as the number of samples (UAV locations where RSRPs are used) increases to localize the BS in the offline mode ($K=50$ in Algorithm~\ref{algorithm:localization}). (b) Localization RMSE after convergence versus the UAV height. }\label{fig:dist_err}
\end{figure}

\begin{figure}[t!]
	\centering
	\includegraphics[width=0.48\textwidth]{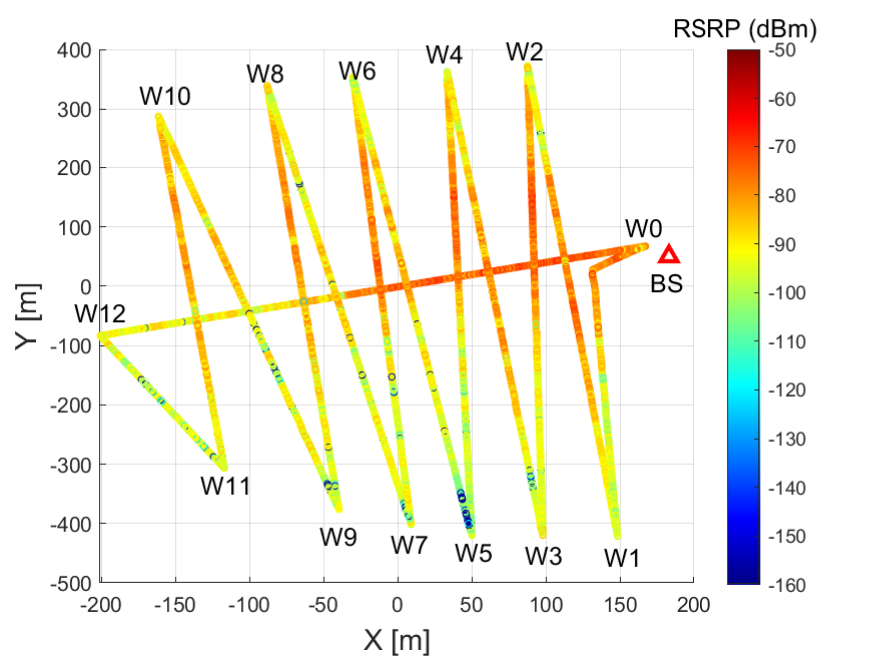}
	\caption{Waypoints and route of the UAV for the online localization where we consider that UAV follows the path $\mathsf{W}_{11}-\mathsf{W}_{10}-\cdots-\mathsf{W}_{0}$.}\label{fig:trajectory_map}
\end{figure}
\begin{figure}[t!]
	\centering
        \subfloat[30~m height]{\includegraphics[width=0.39\textwidth]{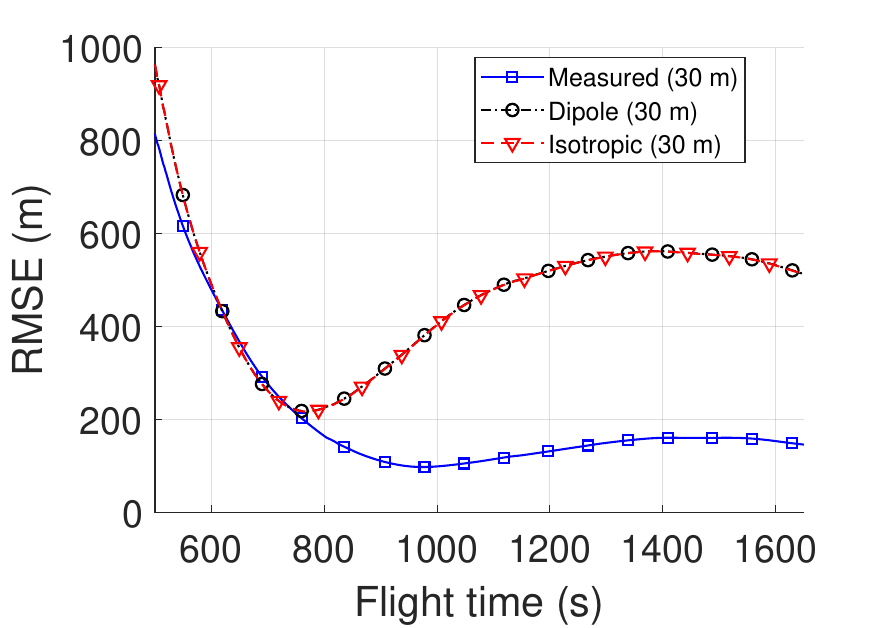}\label{fig:online_30}}
        \vspace{-0.02in}
        \subfloat[70~m height]{\includegraphics[width=0.39\textwidth]{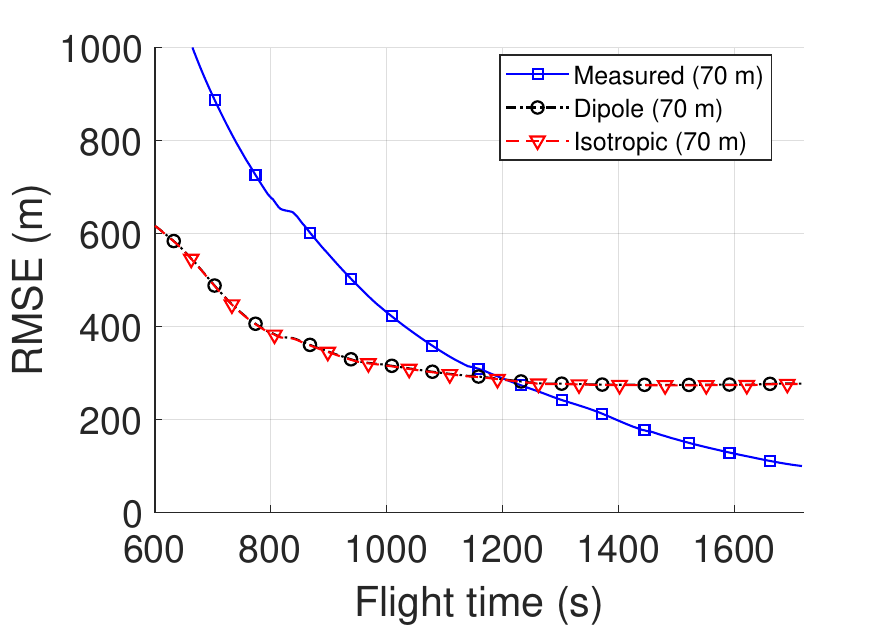}\label{fig:online_70}}
        \vspace{-0.02in}
        \subfloat[90~m height]{\includegraphics[width=0.39\textwidth]{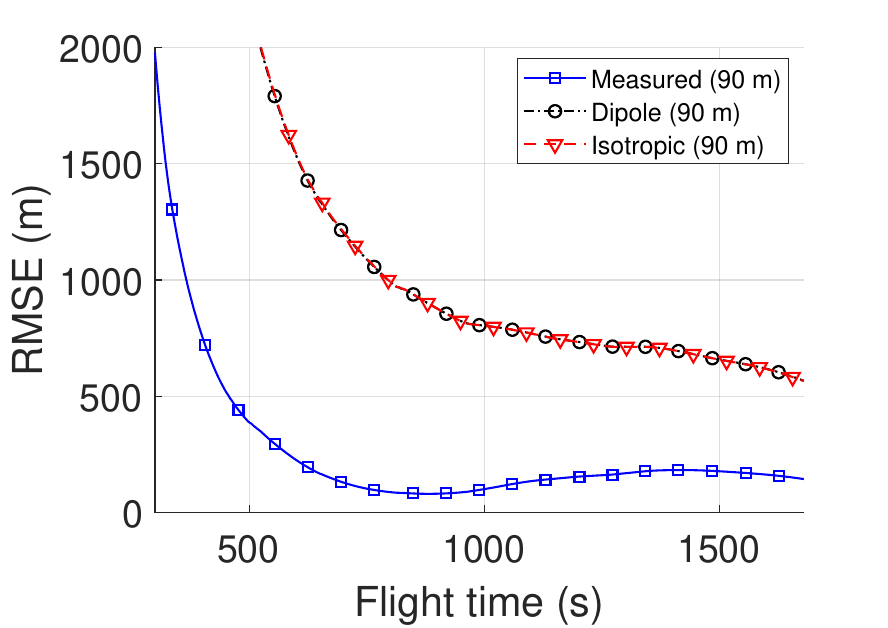}\label{fig:online_90}}
        \vspace{-0.02in}
        \subfloat[110~m height]{\includegraphics[width=0.39\textwidth]{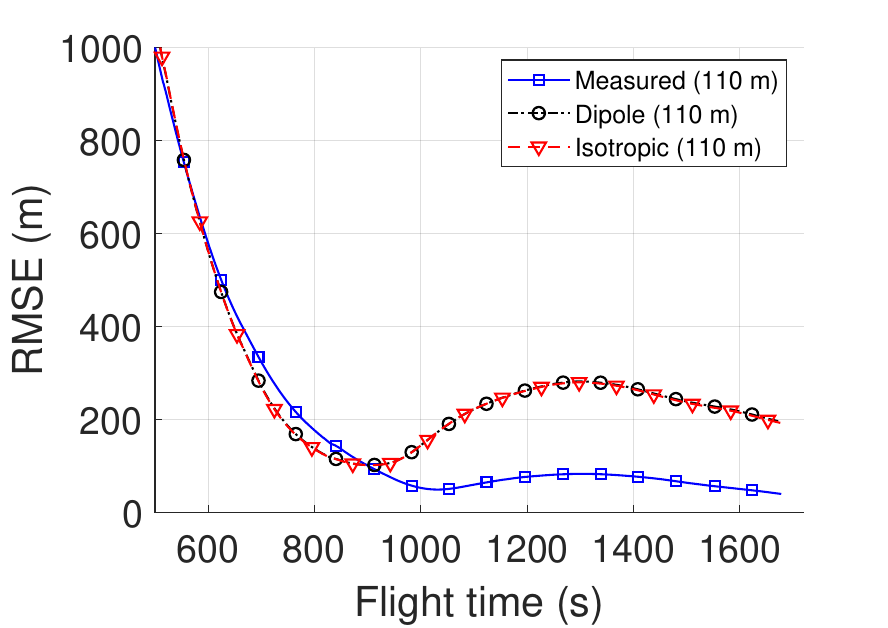}\label{fig:online_110}}
	\caption{The RMSE as a function of flight time in the online localization method using Algorithm~\ref{algorithm:online} with different UAV heights.}\label{fig:online_localization}
\end{figure}

First, in this subsection, the location of the BS tower is estimated by RSRP obtained from offline datasets collected by the UAV. To enhance the accuracy, we initially applied a moving average filter across all RSRP samples to mitigate the impact of rapid fluctuations. Subsequently, we employ Algorithm~\ref{algorithm:localization} for the localization process. We consider the three types of antenna patterns discussed earlier in Section~\ref{sec:3D_ant_rad} (measured, dipole, isotropic). Fig.~\ref{fig:convergence} shows the convergence of the iterative algorithm. The root mean square error (RMSE) is calculated based on the distance error $\sqrt{(\omega^{\rm bs}-\hat{\omega}^{\rm bs})^2\cos(\psi_0)^2+(\psi^{\rm bs}-\hat{\psi}^{\rm bs})^2}\frac{{\rm R}\pi}{180}$ from \eqref{eq:dh}. It is observed that when considering the measured antenna pattern, the distance error is significantly reduced as the number of iterations increases, converging in less than $20$ iterations.  However, the dipole antenna pattern does not improve the distance error. This is due to the fact that the measured antenna pattern accurately captures RSRP and reduces the distance error through iterative updates of the estimated locations. In contrast, the dipole antenna pattern does not improve localization performance as it lacks precision compared to the isotropic antenna pattern. It is worth noting again that we initialize all the algorithms with the isotropic antenna gain at the first step of the iterations.

In our offline localization, we evaluate the RMSE as the number of samples increases for different UAV heights. We randomly choose $N$ samples from the offline collected datasets. In Fig.~\ref{fig:dist_err}, we observed that adopting the measured antenna pattern resulted in higher localization accuracy compared to considering the dipole antenna pattern and isotropic antenna pattern. This indicates that accurate antenna pattern modeling is critical in improving localization performance.

\subsection{Performance of Online RSRP-based Localization}
In this section, we evaluate the performance of BS localization in a real-time scenario using the online mode. During data collection by the UAV, localization is carried out in real-time. Fig.~\ref{fig:trajectory_map} shows the trajectory and waypoints followed by the UAV during the measurement. In the original data collection experiment, the UAV flies away from the BS starting from $\mathsf{W}_{0}$ and proceeding to $\mathsf{W}_{12}$, before returning back to $\mathsf{W}_{0}$ following the path $\mathsf{W}_{0}-\mathsf{W}_{1}-\cdots-\mathsf{W}_{12}-\mathsf{W}_{0}$. On the other hand, the UAV flying toward the BS is a more relevant scenario for online localization. Therefore, we consider the route $\mathsf{W}_{11}-\mathsf{W}_{10}-\cdots-\mathsf{W}_{0}$ and the corresponding RSRP measurements over this trajectory for testing the online localization techniques. 

\begin{figure}[t!]
	\centering       
        \subfloat[30~m height]{\includegraphics[width=0.39\textwidth]{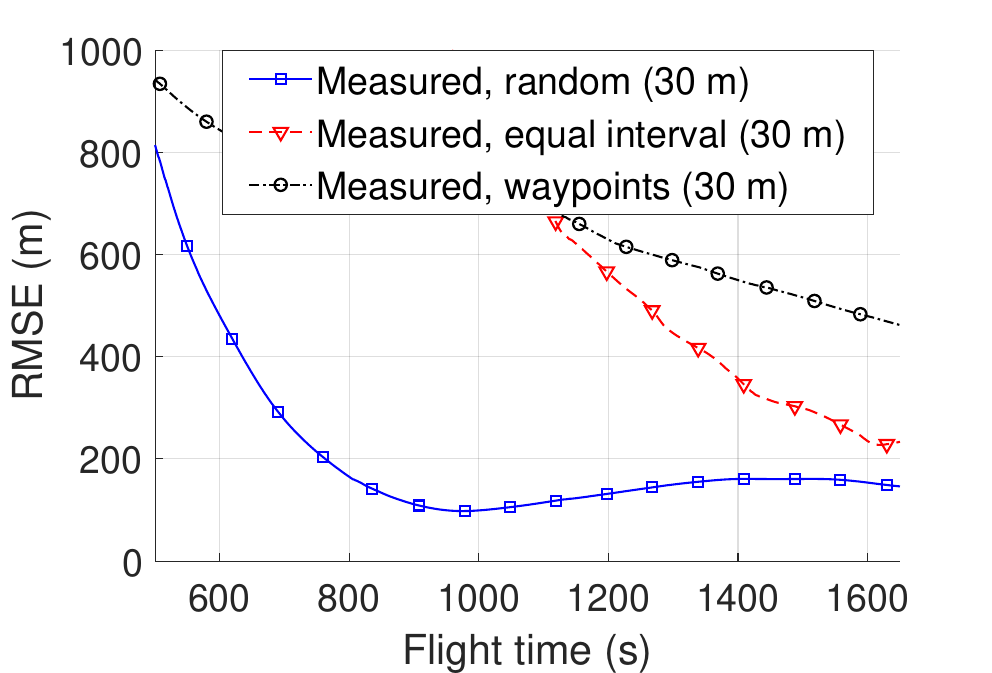}\label{fig:online_mtd_30}}
        \vspace{-0.02in}
        \subfloat[70~m height]{\includegraphics[width=0.39\textwidth]{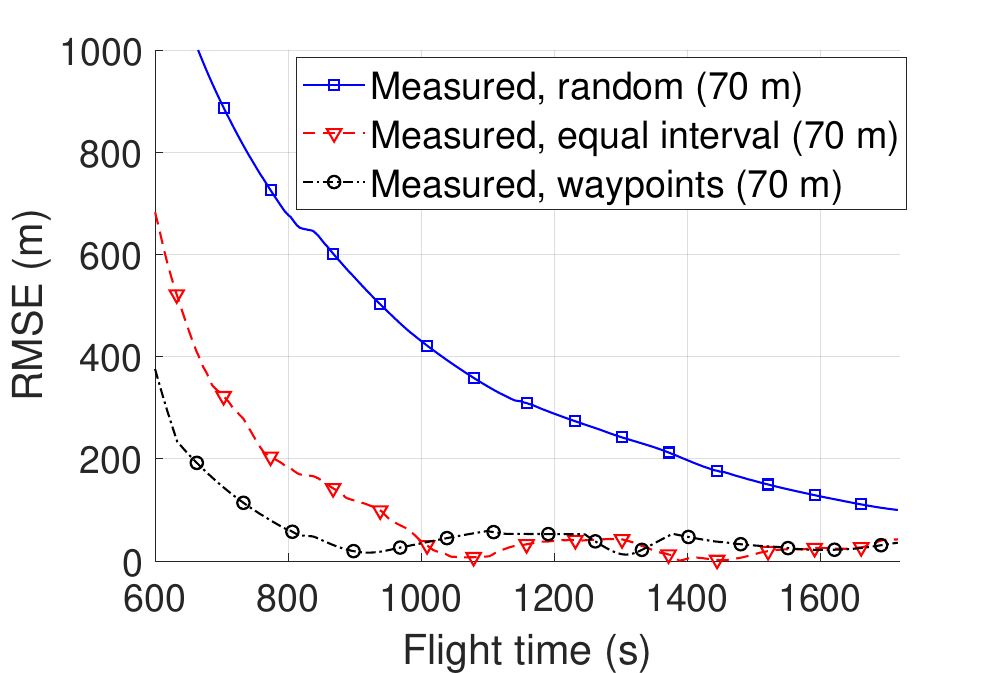}\label{fig:online_mtd_70}}
        \vspace{-0.02in}
        \subfloat[90~m height]{\includegraphics[width=0.39\textwidth]{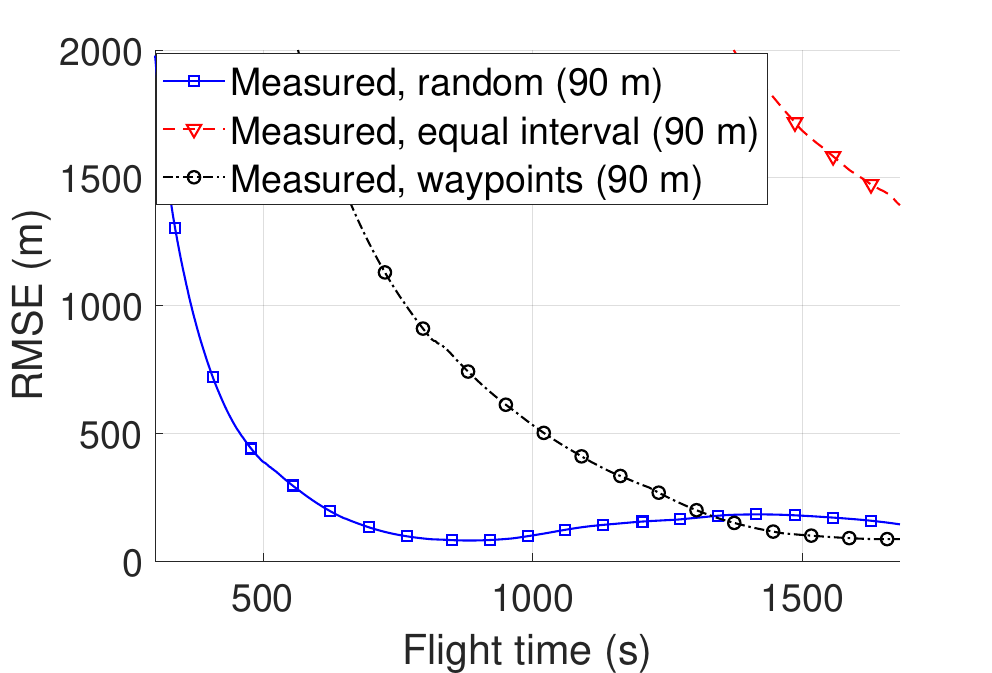}\label{fig:online_mtd_90}}
        \vspace{-0.02in}
        \subfloat[110~m height]{\includegraphics[width=0.39\textwidth]{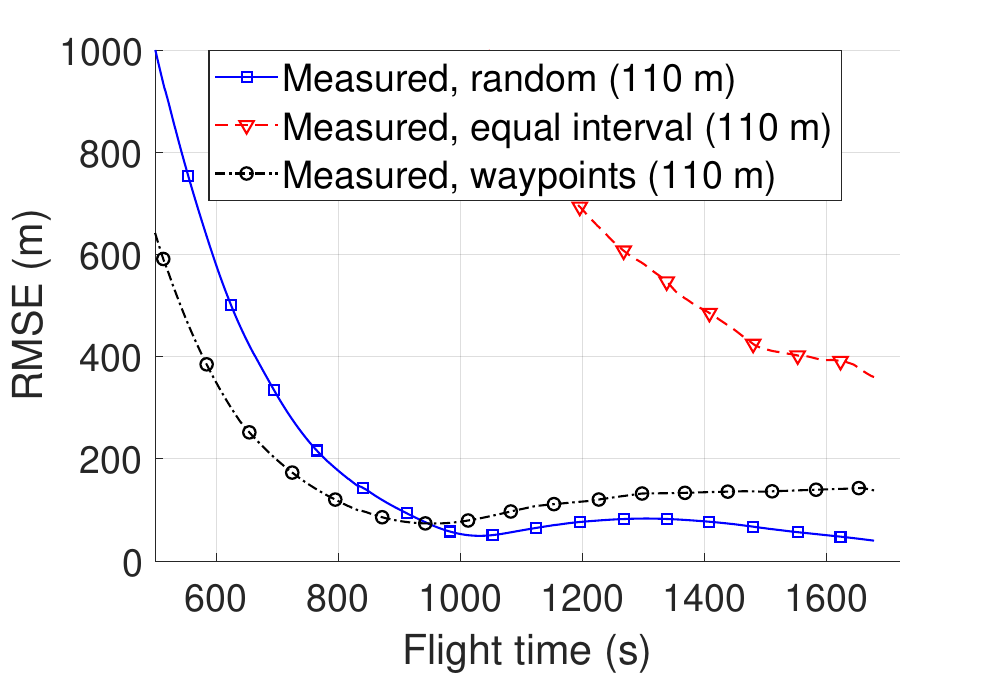}\label{fig:online_mtd_110}}
	\caption{Localization RMSE versus flight time with three different RSRP selection strategies over the trajectory of the UAV, in line 13 of Algorithm~\ref{algorithm:online}, for localizing the signal source.}\label{fig:online_localization_mtd}
\end{figure}

Considering this route, Fig.~\ref{fig:online_30}, Fig.~\ref{fig:online_70}, Fig.~\ref{fig:online_90}, and Fig.~\ref{fig:online_110} show the RMSE as a function of the UAV flight time for different heights. We employ Algorithm~\ref{algorithm:online} for the online localization of the BS, using $M=10$ RSRP measurements from the past locations over the UAV's trajectory, with $N_{\rm max}=10000$. It is observed that in all heights, the RMSE decreases as the flight time increases, and the measured antenna pattern achieves a smaller RMSE compared to the dipole and isotropic antenna patterns at the end of flight time. However, we also observe that early on during the UAV's flight, the dipole and the isotropic antenna patterns outperform the measured antenna pattern at 30~m, 70~m, and 110~m UAV altitudes. This observation holds true until around 750~s for the 30~m UAV height, 1200~s for the 70~m UAV height, and 900~s for the 110~m UAV height.

Fig.~\ref{fig:online_localization_mtd} compares the RMSE of online localization using other alternative approaches to the random selection of the UAV locations where the RSRPs will be used for localization. In particular, on line 13 of Algorithm~\ref{algorithm:online}, we have the option to randomly select $M$ samples from the buffers. However, instead of random selection, we also consider two alternative strategies: equal interval selection and nearby waypoints selection, which were discussed in Section~\ref{sec:online_local}. These strategies introduce temporal and spatial separation among the selected samples, respectively. The results show that the random selection strategy outperforms others at 30~m and 110~m heights, while the nearby waypoints selection strategy achieves the best performance at 70~m and 90~m heights. In addition, the equal interval selection and the nearby waypoints selection strategies perform poorly at 90~m and 110~m heights, and 70~m height, respectively. It implies that the random selection strategy generally yields good performance across all height datasets. On the other hand, while the equal interval selection and nearby waypoints selection strategies can outperform the random selection strategy in certain cases, they may also result in poor performance depending on the specific heights being considered.

\section{Conclusion}\label{sec:conclusion}
In this paper, we study the impact of 3D antenna radiation patterns in air-to-ground path loss modeling. We collect I/Q samples in C-band using an SDR receiver carried by a UAV in a rural environment, to capture signals from an SDR-based LTE BS. The obtained I/Q samples are then post-processed to obtain the RSRP from the target LTE transmission. We model the 3D antenna radiation pattern by using antenna measurements in an anechoic chamber. Then, we can compare the measurement RSRP with the analytical RSRP, obtained using the proposed path loss models that take into account the 3D antenna patterns. We also 
%compare the measured antenna pattern with both constant gain and dipole shape antenna patterns and 
evaluate the accuracy of analytical RSRP considering dipole and isotropic-gain antenna patterns, where we observe significantly better path loss modeling accuracy when the measured antenna patterns are used.  Moreover, we propose an RSRP-based fixed-point iterative localization solution in both offline and real-time online localization scenarios, and we show that the localization accuracy is improved significantly when accurate 3D antenna patterns are used as side information. 
%We observe that the analytical RSRP obtained by the measured antenna pattern is highly close to the measurement RSRP when compared with other antenna pattern models. 
Finally, we estimate the 3D antenna pattern using the measurement RSRP and compare it with the antenna pattern measured at the anechoic chamber. We observe that the directivity of the antenna pattern is similar between the measured and estimated antenna patterns.

\bibliographystyle{IEEEtran}
\bibliography{IEEEabrv,references}
  
\end{document}